\documentclass[10pt,journal,letterpaper,twocolumn,twoside]{IEEEtran}   

\usepackage{amsmath,amsfonts,bm,amssymb,psfrag,ifthen,color,subfigure}
\usepackage{gensymb}
\usepackage{mathrsfs}
\usepackage[justification=centering,font=footnotesize]{caption}
\usepackage{todonotes}
\usepackage{caption}

\allowdisplaybreaks[4]

\usepackage{graphicx}
\usepackage{subfigure}
\usepackage[dvips]{epsfig}
\usepackage[dvips]{graphicx}
\usepackage{amsmath,amsfonts,bm,amssymb,psfrag,ifthen,color,subfigure}
\usepackage[ruled,linesnumbered]{algorithm2e}
\usepackage{ifthen}
\usepackage{setspace}
\usepackage{cite}
\usepackage{url}
\usepackage{longtable}
\usepackage{stfloats}  
\usepackage{graphics,booktabs,color,subfigure}
\usepackage{float}
\usepackage{diagbox}

\usepackage[dvips]{epsfig}
\usepackage[dvips]{graphicx}
\usepackage{amsmath,amsfonts,bm,amssymb,psfrag,ifthen,color,subfigure,amsthm}
\usepackage{setspace}
\usepackage{cite}
\usepackage{url}
\usepackage{longtable}
\usepackage{stfloats}  
\usepackage{graphics,booktabs,color,subfigure}
\usepackage{float}
\usepackage{diagbox}
\usepackage{multirow}

\allowdisplaybreaks[4]

\newcommand{\tabincell}[2]{\begin{tabular}{@{}#1@{}}#2\end{tabular}}

\usepackage{mathrsfs}

\usepackage{graphicx}
\usepackage{makecell}
\usepackage{epstopdf}

\usepackage{graphicx}
\usepackage{todonotes} 

\begin{document}
	
	\title{Semantic Feature Division Multiple Access for
 Digital Semantic Broadcast Channels}
 
	\author{Shuai Ma,~\IEEEmembership{Member,~IEEE}, Zhiye Sun,  Bin Shen, Youlong Wu,~\IEEEmembership{Member,~IEEE}, Hang Li,~\IEEEmembership{Member,~IEEE}, Guangming Shi,~\IEEEmembership{Fellow,~IEEE}, Shiyin Li, and Naofal Al-Dhahir,~\IEEEmembership{Fellow,~IEEE}

 	\thanks{Shuai Ma is with Peng Cheng Laboratory, Shenzhen 518066, China. (e-mail:
 		mash01@pcl.ac.cn).}

		%
	}
	
	\maketitle
	\begin{abstract}

In this paper, we
propose  a digital semantic feature division multiple access (SFDMA) paradigm in
multi-user  broadcast (BC) networks for the inference and the image reconstruction tasks. In this SFDMA scheme,  the multi-user semantic information is encoded  into discrete approximately orthogonal representations, and the encoded semantic features of  multiple users can be simultaneously transmitted in
the same time-frequency resource.
Specifically, for    inference tasks, we design a SFDMA digital BC network based on robust information bottleneck (RIB), which can
achieve a tradeoff between inference performance,
data compression and multi-user interference. Moreover, for image reconstruction tasks,
we develop a SFDMA digital BC network by utilizing a Swin Transformer, which significantly reduces multi-user interference.
More importantly, SFDMA can protect the privacy of users' semantic information, in which
each receiver can only decode its own semantic information.
Furthermore,   we  establish a relationship between  performance  and signal to interference plus noise ratio (SINR), which is fitted by
an   Alpha-Beta-Gamma (ABG) function. Furthermore, an optimal power allocation method is developed for the inference and reconstruction tasks. Extensive simulations verify the effectiveness and superiority of our proposed SFDMA scheme.

	\end{abstract}

	\begin{IEEEkeywords}
	 Sematic broadcast network,  semantic feature division multiple access,   image     reconstruction.
	\end{IEEEkeywords}

	\IEEEpeerreviewmaketitle
\section{Introduction}
\nocite{liu2024near, wu2024deep}
With the  unprecedented increase of various intelligent  applications,  such as massive Internet of Things (IoT), multi-sensory extended reality (XR), autonomous driving, flying vehicles,
    and holographic communication\cite{akyildiz2022metaverse},  it is envisioned that
        the fifth-generation (5G)
wireless communication is
 approaching the Shannon
limits\cite{Strinati_TVT_2019}, and   reaching a resource bottleneck for the massive
connectivity requirements \cite{yang2022ofdm, ihara1978capacity}.
Efficient multiple access schemes are the key solutions  to     increase the     massive connectivity  with high spectral efficiency.

From the first generation (1G) to 5G,
 orthogonal multiple access (OMA)
schemes
schedule  users or groups of users in orthogonal dimensions \cite{clerckx2016rate}, including  frequency division multiple access (FDMA),
time division multiple access (TDMA), code division multiple access (CDMA),   orthogonal frequency division multiple access (OFDMA), and space division multiple access (SDMA), which can   prevent   inter-user interference with  low complexity.
 In addition, the
   non-orthogonal multiple access (NOMA)\cite{Mao2020MultipleAT}, which utilizes the superposition coding (SC) at   transmitters and the
successive interference cancellation (SIC) at   receivers,   has been
   proposed to
    superpose
users in the same time-frequency resource.
Due to the  high SIC complexity, the      number of superposition users in the same time-frequency resource is limited to two \cite{clerckx2021noma}, which restricts the achievable rates.
Thus,   6G needs to explore new  degree of freedom to   satisfy the   massive connectivity \cite{Saad_INW_2020,Zhang_Engineering_2022}.

Fortunately, with the successful development of deep learning (DL), the semantic communication, which
 is inspired from
the human-to-human communication, has the potential to address  the technological challenges in the existing wireless network.
 In   contrast
to the traditional communications, which increase the transmission rate by increasing the physical   resource consumption, such as  bandwidth, antennas and transmit power, semantic communications focus on transmitting only task-related information,  which can significantly reduce the data traffic.
 By exploiting the computing power to alleviate the cost of
 the communication resources \cite{Xie_TSP_2021, Weng_JSAC_2021,ma2023task},    semantic communication
   is
 a promising paradigm shift for the  6G wireless network design.

    In general, there are two typical application tasks of semantic communications, i.e.,   inference tasks and data reconstruction tasks.
     Specifically,
for inference tasks,  semantic communications focus on only transmitting the  information with the specific meanings   for   various tasks \cite{lee2019deep, huang2022toward,hu2023robust,Shao_JSAC_2022, Xie_2023}, instead of accurate data recovery.
For example, a  masked vector quantized-variational auto-encoder (VQ-VAE) method was
proposed in \cite{hu2023robust} to  improve    robustness against semantic noise.
For  downstream inference tasks,   a variational information bottleneck (VIB) \cite{Tishby_arXiv_2000}  based  semantic encoding approach was proposed in \cite{Shao_JSAC_2022} to reduce  the feature transmission latency in
dynamic channel conditions. Furthermore, a   robust VIB based digital    semantic communications
 was designed in \cite{Xie_2023}
that can achieve better inference performance than the baseline methods
with low communication latency.

    For data reconstruction tasks,
    various deep joint source-channel coding (JSCC) semantic communication
schemes were designed for different data modalities, such as  text\cite{Xie_TSP_2021, farsad2018deep ,xie2020lite}, speech\cite{Weng_JSAC_2021, lu2021reinforcement}, image\cite{Xu_TCSVT_2021, kurka2021bandwidth, bourtsoulatze2019deep} and video\cite{Tung_DW_2022, index2017global}.
 More specifically, for text transmission,  a transformer based on  semantic
coding scheme was proposed in \cite{Xie_TSP_2021}  by using sentence similarity
to deal with channel noise and semantic
distortion.
 To accurately recover  speech information  at the semantic
level, a DL-enabled semantic communication system
was designed  in  \cite{Weng_JSAC_2021} by employing a squeeze-and-excitation (SE) network, which is   robust to channel
variations  for   low signal-to-noise (SNR) regions.
 For wireless image transmission, a reinforcement learning based adaptive semantic
coding (RL-ASC) scheme was designed in  \cite{huang2022toward} for  image reconstruction with high semantic similarity and perceptual performance.
  For wireless video conferencing,  an incremental redundancy hybrid automatic repeatrequest
(IR-HARQ) framework was presented in \cite{Jiang2022wireless} that the channel state information (CSI) is considered to allocate the key point
transmission and enhance the performance dramatically.
However, the above existing  semantic communication  works only  focus on the single-user scenarios, which cannot be
 directly applied in the multi-user broadcast communication (BC) network, due to  the single-user semantic
coding schemes cannot effectively deal with the  multi-user interference.

Note that the multi-user semantic  BC networks,  as the most common communication scenario,
 are still  not well studied.
Different from the  single-user semantic communication  scenarios, the multi-user interference is a  critical bottleneck for improving the capacity  of multi-user communication networks.

	Based on the deep neural network (DNN), the authors of \cite{Hu_OTM_2022}  designed a semantic communication system for the broadcast scenario, where the two receivers use semantic recognizer to distinguish	 positive  and negative sentences.
	For visual question answering (VQA) tasks, a multi-user task-oriented communication  system  was developed in \cite{Xie_VQA_2022}  by using multiple antennas linear minimum mean-squared error (L-MMSE) detector and joint source channel decoder, to mitigate the effects of channel distortion and inter-user interference. Moreover, in \cite{Zhang_MUS_2022}, a multi-user semantic communication system is studied to execute object-identification tasks, where correlated source data among different users is transmitted via a shared channel.
By combining the similar components of the extracted semantic features,
  a model division multiple access (MDMA) scheme was proposed in \cite{zhang2023model},
  in which the common semantic
information of multi-user is transmitted within the same
time-frequency resources and the  personalized semantic
information is transmitted separately.
	In \cite{Luo_MM_2022},  the authors proposed a  multi-modal information fusion scheme for multi-user semantic communications,  where the wireless channel acts as a medium to fuse multi-modal data where a receiver   retrieves semantic information without the need to perform multiuser signal detection.
By using attention and residual structure
modules, a DL-based multiple
access method was designed  in \cite{Zhang_TCCN_2023}  for continuous semantic symbols analog transmission
in image reconstruction tasks.

In the above existing  works\cite{Hu_OTM_2022,Xie_VQA_2022,Zhang_MUS_2022,zhang2023model,Luo_MM_2022,Zhang_TCCN_2023}, JSCC  directly maps
the source data into continuous channel input symbols, which is incompatible
  with   current digital communication systems.
Moreover,  the direct transmission of continuous
feature representations requires analog
modulation or a full-resolution constellation, which brings
huge burdens for resource-constrained radio frequency
systems.
In addition, the performance of semantic communication depends on     transmission power. The existing semantic communication performance measurements  are  end-to-end, such as  classification accuracy for inference tasks, which  has no consideration of transmission power. The main reason is that  DL based semantic encoders  are  generally highly complex nonlinear functions, and it is hard to derive analytical relationships between end-to-end performance measures related to the transmit power. Therefore,   there is no theoretical basis for  adaptive power control for  semantic communications, which leads to performance degradation   in random fading channels.

In short, mitigating the multi-user interference and missing theoretical model for adaptive power allocation in the semantic communication are the two major research challenges. In this paper, we explore a new resource domain—semantic feature domain, and propose a semantic feature division multiple access (SFDMA) scheme,  where the information of multiple users are encoded  in the distinguishable feature subspaces.
Specifically, in proposed SFDMA scheme,  the encoded semantic features of  multiple users  are approximately   orthogonal to each other, and can be simultaneously transmitted in
the same time-frequency resource. This scheme is applied in two scenarios: one focused on inference tasks and the other on image reconstruction.
 The main contributions of this paper can be summarized as follows:
\begin{itemize}
	\item 	
	By exploring the feature domain,  we propose a SFDMA  scheme for multi-user  BC networks, in which the base station (BS)  encodes   the multi-user information into discrete semantic feature representations, ensuring  semantic features   of   different users are   approximately orthogonal, which effectively alleviates multi-user interference.

	\item  For   inference tasks, we design a  SFDMA   BC network based on a robust information bottleneck (RIB), which can
	  achieve a tradeoff between   inference performance,
data compression and multi-user interference.
Specifically, by exploiting RIB,   the SFDMA   BC network  maximizes the semantic coded redundancy for robust  transmission, while restricting the multi-user  interference and  extract sufficient semantic information for the   inference tasks. Due to the computational intractability of mutual information, we derive the tractable variational upper bound of the RIB   by utilizing the variational approximation technique.  The proposed RIB based SFDMA scheme can  achieve efficient   transmission of  the  semantic information, while protecting user semantic information from being decoded by other users.
	
	\item 	   For   image  reconstruction tasks, we design a  SFDMA   BC network based on the   Swin Transformer, a novel and effective computer vision model\cite{liu2021swin}.
The network broadcasts discrete multi-user information at the same time and frequency resource.
	Moreover, our proposed SFDMA scheme protects the privacy of users' semantic information    being decoded  by other users.

\item  Furthermore, we  establish  a relationship between performance  and signal to interference plus noise ratio (SINR), which we approximately fit to an Alpha-Beta-Gamma (ABG) formula. To the best of our knowledge, this is the first analytical expression between performance  and SINR for semantic BC networks.
    Based on the ABG function,  we propose an optimal power allocation method for the semantic BC network with  inference and reconstruction tasks. Our proposed  power allocation method   can   effectively guarantee the quality of   service (QoS)  semantic communication in random fading channels.

\end{itemize}

The rest of this paper is organized as follows.  Section II and III present the
  SFDMA   BC network with inference and  image reconstruction  tasks, respectively.
 Section IV presents the optimal power allocation scheme for the SFDMA BC network.
  Then,  the extensive simulation
results are presented in Section V, followed by the concluding
remarks in Section VI.

\emph{Notations}:  Vectors and matrices are represented by boldfaced lowercase and uppercase letters, respectively.
The symbols ${\mathbb{E}}\left\{ \cdot \right\}$,
${\left(  \cdot  \right)^{{T}}}$,  represent   the expectation, transpose, respectively.
${{I}}\left( {X;Y} \right)$ denotes the mutual information between input $X$ and output $Y$,
 and ${\bf{I}}$ is an identity matrix.
\begin{table}[htbp]
\caption{Summary of Key Notations}
\label{tablepar}
\centering
\begin{tabular}{|m{1.8cm}<{\centering}|m{5.8cm}|}
	\hline
	\rule{0pt}{7pt}Notations  &    Meanings \\ \hline
	
	\rule{0pt}{6.5pt} ${{\mathbf{s}}_i}$ &  {The input data of User $i$ } \\ \hline
	
	\rule{0pt}{6.5pt} ${{\mathbf{u}}_i}$ &  {The semantic information of User $i$} \\ \hline
	
	\rule{0pt}{6.5pt} ${{\mathbf{a}}_i}$ &  {The extracted semantic features vectors of User $i$} \\ \hline
	
	\rule{0pt}{6.5pt} ${{\mathbf{z}}_i} $  &  {Quantized signal of User $i$}  \\ \hline

    \rule{0pt}{6.5pt} ${{\mathbf{x}}_i} $  &  {Modulation signal of User $i$}  \\ \hline
	
    \rule{0pt}{6.5pt} ${f_{{\varphi _i}}}\left( \cdot \right) $  &  {The semantic encoder and its parameters $\varphi_i$ at the BS of User $i$}  \\ \hline
    \rule{0pt}{6.5pt} ${\Psi _i}\left( \cdot \right)$  &  {Quantizer of User $i$}  \\ \hline

    \rule{0pt}{6.5pt} $\Theta _{\rm{m}}^i\left( {\cdot} \right) $  &  {Modulation of User $i$}  \\ \hline

    \rule{0pt}{6.5pt} ${f_{{\psi _i}}}\left( \cdot \right)$  &  {The encoder and its parameters $\psi_i$ of User $i$}  \\ \hline

    \rule{0pt}{6.5pt} ${f_{{\theta _i}}}\left( \cdot \right)$  &  {The  decoder and its parameters $\theta_i$ of User $i$}  \\ \hline

    \rule{0pt}{6.5pt} ${{\mathbf{y}}_i} $  &  {The signal received by User $i$}  \\ \hline
    \rule{0pt}{6.5pt} ${\widehat {\bf{u}}_i} $  &  {Estimated semantic information of User $i$}  \\ \hline
	\rule{0pt}{6.5pt}$d $ &  \tabincell{c}{Dimension of feature vector}\\ \hline
\end{tabular}
\end{table}

\begin{figure}[h]
	\centering
	\includegraphics[width=9cm]{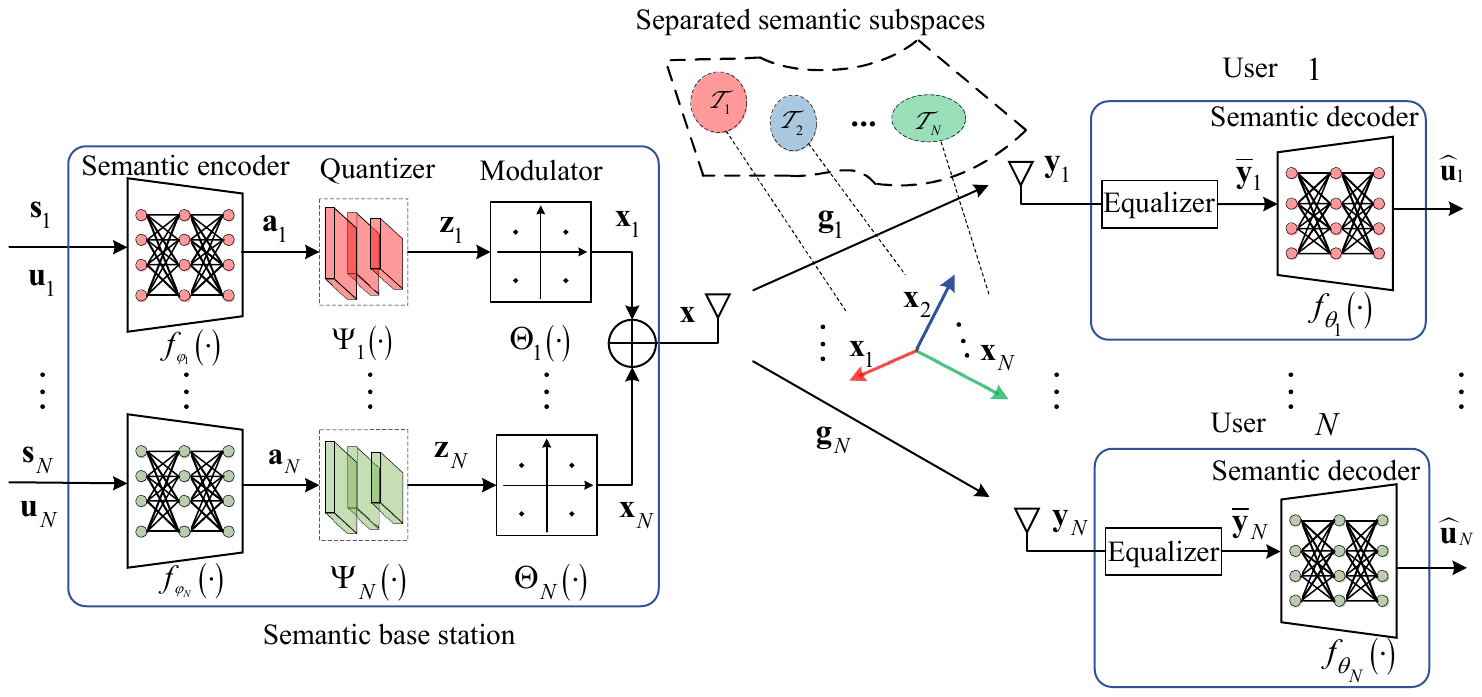}
	\caption{A digital SFDMA BC network model}
	\label{system} 
\end{figure}
\section{ SFDMA for Task-oriented Semantic BC Network}

In this section, we present the SFDMA for task-oriented semantic BC network, and in the section, we devote three subsections to an overview of the task-oriented network model, SFDMA and RIB.

\subsection{ Multi-user semantic BC network}

Considering a   task-oriented   multi-user semantic BC network, as depicted in Fig.~\ref{system}, where a BS aims to transmit the data $\left\{ {{{\bf{s}}_i}} \right\}_{i = 1}^N$ with implicit semantic information  $\left\{ {{{\bf{u}}_i}} \right\}_{i = 1}^N $  to $N$ users simultaneously for inference tasks, and ${{\bf{u}}_i}$ denotes the   intended semantic information of User $i$, where $i = 1,...,N$.
Note that in the structure, the encoder ${f_{{\psi _i}}}$ is the neural network used for the entire encoding process. In the encoder, the semantic encoder ${f_{{\varphi _i}}}$ is the first module in the encoder used to extract analog semantic features.

 To extract the semantic information ${{\bf{u}}_i}$,   we design  a semantic encoder network  ${f_{\varphi_i} }\left( \cdot \right)$, which  extracts semantic feature  $\mathbf{a}_i$ from the source data $\mathbf{s}_i$  as follows
\begin{align}\label{000}
{{\bf{a}}_i} = {f_{{\varphi _i}}}\left( {\bf{s}}_i \right),~i\in\{1,..,N\}.
\end{align}

Note that, the  feature representations $\left\{{\mathbf{a}_i}\right\}_{i=1}^N$  are continuous, and
 the direct transmission of
continuous feature representation  needs to be modulated with
analog modulation or a full-resolution constellation, which
brings huge burdens for resource-constrained transmitter and
poses implementation challenges on the current radio frequency
(RF) systems.  Thus,
 the continuous  feature representations $\mathbf{a}_i$  are quantized    as follows
\begin{align}
\mathbf{z}_i={\Psi _i}\left( {{{\bf{a}}_i}} \right),~i\in\{1,..,N\},\label{quantizer}
\end{align}
where ${\Psi _i}\left(  \cdot  \right)$ denotes the quantizer for  $\mathbf{a}_i$.
Specifically, this quantizer transforms the real-valued input into a discrete set of values, typically +1 and -1, according to the sign of the input. This deterministic binarization allows the network to perform efficient binary operations while preserving the ability to propagate gradients using straight-through estimator (STE) \cite{courbariaux2016binarized}.
Then, the quantized signal $\mathbf{z}_i$   is modulated and power normalized, i.e.,
\begin{align}\label{000}
{{\bf{x}}_i} = {\Theta _i}\left( {{{\bf{z}}_i}} \right),~i\in\{1,..,N\},
\end{align}
where ${\Theta _i}\left(  \cdot  \right)$ denotes the modulator and power normalizer for $\mathbf{z}_i$. In this work, binary phase shift keying (BPSK) modulation is employed.

  Then, by applying power amplification, the  transmitter broadcasts the semantic encoding information $\left\{ {{{\bf{x}}_i}} \right\}_{i = 1}^N $ to the $N$ users.
    The received signal of User $i$ is given as
   \begin{align}\label{00}
{{\bf{y}}_i} = \underbrace {{\bf{g}}_i{\sqrt {{p_i}}\odot{\bf{x}}_i}}_{{\rm{desired\:singal}}} + \underbrace {{\bf{g}}_i\sum\limits_{j = 1,j \ne i}^N {{\sqrt {{p_j}} \odot{\bf{x}}_j}} }_{{\rm{interference}}} + \underbrace {{{\bf{n}}_i}}_{{\rm{noise}}},~i\in\{1,..,N\},
 \end{align}
 where  $ {\bf{g}}_i$ is channel gain  between BS and User $i$, $p_i$ is the allocated  transmitted power for   the signal ${\bf{x}}_i$, and ${{\bf{n}}_i} \sim \mathcal{CN}\left( {0,\sigma _i^2{\bf{I}}} \right)$ denotes the received additive white Gaussian noise of User $i$,
  $ {{\bf{g}}}_i{{\bf{x}}_i}$ is the desired signal of  User $i$, ${\bf{g}}_i\sum\limits_{j = 1,j \ne i}^N {{{\bf{x}}_j}} $ is the multi-user semantic interference.

By applying the equalizer, the equalized received  signal   ${\overline {\bf{y}} _i}$ is given as
\begin{align}
{\overline {\bf{y}} _i} = \sqrt {{p_i}} {{\bf{x}}_i} + \sum\limits_{j = 1,j \ne i}^N {\sqrt {{p_j}} {{\bf{x}}_j}}  + \frac{{{{\bf{n}}_i}}}{{{\bf{g}}_i}},i \in \{ 1,..,N\}, \label{received_signal}
\end{align}
where  ${\overline {\bf{y}}_i}$  is decoded by ${f_{{\theta _i}}}\left(\cdot  \right)$
\begin{align}\label{eq_received_signal}
{\widehat {\bf{u}}_i} = {f_{{\theta _i}}}\left( {\overline {\bf{y}} _i} \right),~i\in\{1,..,N\},
\end{align}
where ${\bf{u}}_i$ is the inference information of ${{u}}_i$.

\subsection {SFDMA  for BC   networks}

	Note that, in contrast to   the single user point-to-point semantic communication without user interference, in  multi-user  semantic BC networks, the BS  sends the semantic information of multiple users to each user at the same time-frequency resources, and  multi-user interference is a key bottleneck  that restricts  the transmission rate and QoS.

	To overcome this bottleneck, we propose a SFDMA scheme for semantic
	BC networks. Specifically, via the encoder ${f_\psi }\left( \cdot \right)$, the   multi-user data $\left\{ {{{\bf{s}}_i}} \right\}_{i = 1}^N$ are encoded  to discrete semantic features $\left\{ {{{\bf{x}}_i}} \right\}_{i = 1}^N $, i.e.,
	\begin{align}
		\left\{ {{{\bf{x}}_1}{\rm{,}}...{\rm{,}}{{\bf{x}}_N}} \right\} = {f_\psi }\left( {{{\bf{s}}_1}{\rm{,}}...{\rm{,}}{{\bf{s}}_N}{\rm{,}}{{\bf{g}}_1}{\rm{,}}...{\rm{,}}{{\bf{g}}_N}} \right){\rm{,}}
	\end{align}
	where ${\bf{g}}_i $ denotes the channel gain from the BS to the user $i$, $i \in \left\{ {1,...,N} \right\}$.
	
				In the SFDMA scheme, the semantic features of multi-user are   approximately orthogonal, i.e.,
	\begin{align}\frac{{{\bf{x}}_i^H{{\bf{x}}_j}}}{{\left\| {{{\bf{x}}_i}} \right\|\left\| {{{\bf{x}}_j}} \right\|}} \to 0,\forall i \ne j.
	\end{align}
	
	Since the semantic encoded vectors of users are mutually orthogonal, the semantic information of multiple users can be transmitted simultaneously on  same time-frequency resources, avoiding interference among users. In addition, SFDMA can achieve a certain degree of semantic information privacy protection, that is, users can only decode their own semantic information and cannot decode the semantic information of other users.

\subsection {RIB principle}
For inference tasks, the more semantic data is transmitted to the users, the more accurate inference can be attained.
 However, more data transmission  will increase communication overhead and interference to other users. As a result, there is an inherent trade-off among inference performance, data compression and multi-user interference. Such a trade-off is a key factor in inference task-oriented communication design.
To achieve the above trade-off,  we formulate SFDMA BC   for inference tasks   problem based on RIB principle, which concludes three mutual information terms. Note that the mutual information quantifies the amount of information shared between two random variables, indicating the extent of their dependency. Its physical significance lies in measuring the reduction in uncertainty of one variable upon knowing the other, specifically:
\begin{align}\label{IB_Obj}
\mathop {\min }\limits_{\left\{ {{p_{{\psi _i}}}\left( {{{\bf{x}}_i}|{{\bf{s}}_i}} \right)} \right\}_{i = 1}^N} \sum\limits_{i = 1}^N { - I\left( {{U_i};{Y_i}} \right) - {\omega _i}\left[ {I\left( {X;{Y_i}} \right) - I\left( {{S_i};{Y_i}} \right)} \right]} ,
\end{align}
where
  $ \omega_i > 0$  is a weighted  parameter,
${I\left( {{U_i};{Y_i}} \right)}$ denotes the mutual information between the
	received signal ${{Y_i}}$ and the semantic information ${{U_i}}$, higher mutual information indicates better inference performance,    ${I\left( {{X};{Y_i}} \right)}$ denotes the mutual information between the
received signal ${{Y_i}}$ and the transmitted signal ${{X}}$, ${I\left( {{S_i};{Y_i}} \right)}$
denotes the mutual information between the
received signal ${{Y_i}}$ and the source signal ${{S_i}}$. The term  $I\left( {X;{Y_i}} \right) - I\left( {{S_i};{Y_i}} \right)$ represents the redundant information in the transmitted signal compared to the source signal, which helps to overcome noise and multi-user interference.

Let ${{\cal L}_{\rm{RIB}}}$ denote the objective function of the optimization problem \eqref{IB_Obj}, which   can be equivalently reformulated as
 \begin{align}\label{9}
{{\cal L}_{\rm{RIB}}} &=
 \sum\limits_{i = 1}^N {{\mathbb{E}_{p\left( {{{\bf{s}}_i},{{\bf{u}}_i}} \right)}}\left\{ {{\mathbb{E}_{{p_{{\psi _i}}}\left( {{{\bf{y}}_i}|{{\bf{s}}_i}} \right)}}\left[ { - \log {p_{{\psi _i}}}\left( {{{\bf{u}}_i}|{{\bf{y}}_i}} \right)} \right]} \right.} \nonumber\\
 &\left. { + {\omega _i}{\mathbb{E}_{{p_{{\psi _i}}}\left( {{\bf{x}}|{{\bf{s}}_i}} \right)}}\left[ {H\left( {{Y_i}|{\bf{x}}} \right)} \right] - {\omega _i}{H_{{\psi _i}}}\left( {{Y_i}|{{\bf{s}}_i}} \right) - H\left( {{U_i}} \right)} \right\}\nonumber\\
 & = \sum\limits_{i = 1}^N {{\mathbb{E}_{p\left( {{{\bf{s}}_i},{{\bf{u}}_i}} \right)}}\left\{ {{\mathbb{E}_{{p_{{\psi _i}}}\left( {{{\bf{y}}_i}|{{\bf{s}}_i}} \right)}}\left[ { - \log {p_{{\psi _i}}}\left( {{{\bf{u}}_i}|{{\bf{y}}_i}} \right)} \right]} \right.} \nonumber\\
 &\left. { + {\omega _i}{\mathbb{E}_{{p_{{\psi _i}}}\left( {{\bf{x}}|{{\bf{s}}_i}} \right)}}\left[ {H\left( {{Y_i}|{\bf{x}}} \right)} \right] - {\omega _i}{H_{{\psi _i}}}\left( {{Y_i}|{{\bf{s}}_i}} \right)} \right\},
\end{align}
where  ${H\left( {{U_i}} \right)}$ is ignored in \eqref{9} since it is  constant.

However, the calculation of the posterior ${p_{{\psi _i}}}\left( {{{\bf{u}}_i}|{{\bf{y}}_i}} \right)$ in   \eqref{9} involves high-dimensional integrals, i.e.,
\begin{align}\label{10}
{p_{{\psi _i}}}\left( {{{\bf{u}}_i}|{{\bf{y}}_i}} \right) &= \frac{{\int {{p_{{\psi _i}}}\left( {{{\bf{s}}_i},{{\bf{u}}_i}} \right){p_{{\psi _i}}}\left( {{{\bf{y}}_i}\left| {{{\bf{s}}_i}} \right.} \right)} d{{\bf{s}}_i}}}{{{p_{{\psi _i}}}\left( {{{\bf{y}}_i}} \right)}}\nonumber\\
& = \frac{{\int {{p_{{\psi _i}}}\left( {{{\bf{s}}_i},{{\bf{u}}_i}} \right){p_{{\psi _i}}}\left( {{{\bf{y}}_i}\left| {{{\bf{s}}_i}} \right.} \right)} d{{\bf{s}}_i}}}{{\int {{p_{{\psi _i}}}\left( {{{\bf{s}}_i},{{\bf{u}}_i}} \right)} {p_{{\psi _i}}}\left( {{{\bf{y}}_i}\left| {{{\bf{u}}_i}} \right.} \right)d{{\bf{u}}_i}d{{\bf{s}}_i}}}.
\end{align}

To address this challenge, we exploit variational distributions ${q_{{\theta _i}}}\left( {{{\bf{y}}_i}} \right)$  and ${q_{{\theta _i}}}\left( {{{\bf{u}}_i}\left| {{{\bf{y}}_i}} \right.} \right)$  to approximate the   distributions $p\left( {{{\bf{y}}_i}} \right)$ and $p\left( {{{\bf{u}}_i}\left| {{{\bf{y}}_i}} \right.} \right)$, respectively \cite{alemi2016deep}, where ${\theta _i}$ is the parameter of semantic decoder of User $i$, and is used to compute the inference result ${\widehat {\bf{u}}_i}$.
Specifically, the upper bound of the   first term
${\mathbb{E}_{p\left( {{{\bf{s}}_i},{{\bf{u}}_i}} \right)}}\left\{ {{\mathbb{E}_{{P_{{\psi _i}}}\left( {{{\bf{y}}_i}|{{\bf{s}}_i}} \right)}}\left[ { - \log {p_{{\psi _i}}}\left( {{{\bf{u}}_i}|{{\bf{y}}_i}} \right)} \right]} \right\} $ of  \eqref{9} is given as
\begin{subequations}
\begin{align}
&{\mathbb{E}_{p\left( {{{\bf{s}}_i},{{\bf{u}}_i}} \right)}}\left\{ {{\mathbb{E}_{{p_{{\psi _i}}}\left( {{{\bf{y}}_i}|{{\bf{s}}_i}} \right)}}\left[ { - \log {p_{{\psi _i}}}\left( {{{\bf{u}}_i}|{{\bf{y}}_i}} \right)} \right]} \right\}\label{main}\nonumber\\
 &   = {\mathbb{E}_{p\left( {{{\bf{s}}_i},{{\bf{u}}_i}} \right)}}\left\{ {{\mathbb{E}_{{p_{{\psi _i}}}\left( {{{\bf{y}}_i}|{{\bf{s}}_i}} \right)}}\left[ { - \log {q_{{\theta _i}}}\left( {{{\bf{u}}_i}|{{\bf{y}}_i}} \right)} \right]} \right.\nonumber\\
 & \qquad \quad - \left. {{D_{{\rm{KL}}}}\left( {{p_{{\psi _i}}}\left( {{{\bf{s}}_i}\left| {{{\bf{y}}_i}} \right.} \right)\left\| {{q_{{\theta _i}}}\left( {{{\bf{s}}_i}\left| {{{\bf{y}}_i}} \right.} \right)} \right.} \right)} \right\}\\
 &  \le {\mathbb{E}_{p\left( {{{\bf{s}}_i},{{\bf{u}}_i}} \right)}}\left\{ {{\mathbb{E}_{{p_{{\psi _i}}}\left( {{{\bf{y}}_i}|{{\bf{s}}_i}} \right)}}\left[ { - \log {q_{{\theta _i}}}\left( {{{\bf{u}}_i}|{{\bf{y}}_i}} \right)} \right]} \right\}\label{sub},
\end{align}
\end{subequations}
where ${D_{{\rm{KL}}}}$ represents the Kullback-Leibler divergence, which is a measure of the difference between two probability distributions, thus the inequality \eqref{sub} holds due to ${D_{{\rm{KL}}}} \ge 0$.

Moreover, since ${{\bf{y}}_{i}}$ is  corrupted by channel noise and multi-user interference, we can derive a lower bound for the entropy
term ${H_{{\psi _i}}}\left( {{Y_i}|{{\bf{s}}_i}} \right)$ as
\begin{align}\label{8}
{H_{{\psi _i}}}\left( {{Y_i}|{{\bf{s}}_i}} \right) \ge {H_{{\psi _i}}}\left( {{X_i}|{{\bf{s}}_i}} \right).
\end{align}

Hence, the objective function in \eqref{9} is upper bounded by
\begin{align}\label{VRIB}
{{\cal L}_{\rm{RIB}}} &\le \sum\limits_{i = 1}^N {{\mathbb{E}_{p\left( {{{\bf{s}}_i},{{\bf{u}}_i}} \right)}}\left\{ {{\mathbb{E}_{{P_{{\psi _i}}}\left( {{{\bf{y}}_i}|{{\bf{s}}_i}} \right)}}\left[ { - \log {q_{{\theta _i}}}\left( {{{\bf{u}}_i}|{{\bf{y}}_i}} \right)} \right]} \right.}\nonumber \\
&~~\left. { + {\omega _i}{\mathbb{E}_{{p_{{\psi _i}}}\left( {{\bf{x}}|{{\bf{s}}_i}} \right)}}\left[ {H\left( {{Y_i}|{\bf{x}}} \right)} \right] - {\omega _i}{H_{{\psi _i}}}\left( {{Y_i}|{{\bf{s}}_i}} \right)} \right\}
\end{align}

The conditional entropy terms   $ {{H_{{\psi _i}}}\left( {{Y_i}|{{\bf{s}}_i}} \right)}$ and $ {H\left( {{Y_i}|{\bf{x}}} \right)}$ in \eqref{VRIB} are analytically computable with respect to the parameters ${\psi _i}$. The DNN-based encoder is defined by the factorial distribution${p_{{\psi _i}}}\left( {{{\bf{y}}_i}|{{\bf{s}}_i}} \right) = \prod\nolimits_{j = 1}^d {{p_{{\psi _i}}}\left( {{y_{i,j}}\left| {{{\bf{s}}_i}} \right.} \right)} $. Specifically, the entropy terms can be decomposed into the following summations:
\begin{subequations}
	\begin{align}
	{H_{{\psi _i}}}\left( {{Y_i}|{{\bf{s}}_i}} \right) = \sum\limits_{j = 1}^d {{H_{{\psi _i}}}\left( {{Y_{i,j}}|{{\bf{s}}_{i}}} \right)}, \\
	H\left( {{Y_i}|{\bf{x}}} \right) = \sum\limits_{j = 1}^d {H_j\left( {{Y_{i,j}}|{{\bf{x}}_j}} \right)}.
	\end{align}
\end{subequations}

By further applying Monte Carlo sampling, we are able to obtain an unbiased estimate of the gradient and hence optimize the objective using stochastic gradient descent.  We have the following empirical estimation:
\begin{align}
&{{\cal L}_{\rm{RIB}}} \le\sum\limits_{i = 1}^N {\frac{1}{M}\sum\limits_{m = 1}^M {\left\{ {\frac{1}{L}\sum\limits_{l = 1}^L {\left[ { - \log {q_{{\theta _i}}}\left( {{{\bf{u}}_i}^{\left( m \right)}|{{\bf{y}}_i}^{\left( {m,l} \right)}} \right)} \right.} } \right.} } \nonumber\\
&{ + {\omega _i}\sum\limits_{j = 1}^d {\left. {H\left( {{Y_{i,}}_j|{{\bf{x}}_j}^{\left( {m,l} \right)}} \right)} \right]} }{ - {\omega _i}\sum\limits_{j = i}^d {{H_{{\psi _i}}}\left( {{Y_{i,}}_j|{{\bf{s}}_i}^{\left( m \right)}} \right)} }\Bigg \}, \label{VRIB2}
\end{align}
where $M$ is the batch size of training dataset, $L$ is the sample times for each pair $\left\{ {\left( {{\bf{s}}_{_i}^{(m)},{\bf{u}}_{_i}^{(m)}} \right)} \right\}_{{\rm{m}} = 1}^M$, and $d$ represents the dimension of signal vector $Y$ and $X$.

The first term $- \log {q_{{\theta _i}}}\left( {{{\bf{u}}_i}^{\left( m \right)}|{{\bf{y}}_i}^{\left( {m,l} \right)}} \right)$, is maximized, which is equivalent to minimizing the cross-entropy between $\bf{u}_i$ and $\hat{\bf{u}}_i$.
The second term ${H\left( {{Y_{i,}}_j|{{\bf{x}}_j}^{\left( {m,l} \right)}} \right)}$ can be computed for different discrete value of ${\bf{x}}_j$, the conditional distribution $p \left( {\bf{y}}_i | {\bf{x}} \right)$ depends on the allocated transmitted power $p_i$, where $i=1,2,...,N$, and the noise variance $\sigma^2$, as shown in \eqref{received_signal}.
The third term ${{H_{{\psi _i}}}\left( {{Y_{i,}}_j|{s_i}^{\left( m \right)}} \right)}$ is determined by $p \left( {\bf{y}}_{i,j} | {\bf{s}}_i \right)$. Specifically, we model the conditional distribution $p_{\varphi}\left( {\bf{a}}_{i,j} | {\bf{s}}_i \right)$ as a multivariate Gaussian distribution due to the feature of neural network, combined with the binarized quantification method and the BPSK modulation, the conditional distribution $p_\psi\left( {\bf{x}}_{i,j} | {\bf{s}}_i \right)$ is modeled as a uniform Bernoulli distribution, then $p\left( {\bf{y}}_{i,j} | {\bf{s}}_i \right)$ is depicted as:


\vspace{-0.2cm}
\begin{align}
	{p_\psi }({{\bf{y}}_{i,j}}|{\bf{s}}) = &\sum\limits_{t = 1}^T {\frac{1}{T}} {f_\varepsilon }(\sqrt {{p_i}} {{\bf{x}}_{i,j}} + \sum\limits_{c = 1,c \ne i}^N {\sqrt {{{p}_c}} {{\bf{x}}_{c,j}}}  \\
	& + \frac{{{{\bf{n}}_{i,j}}}}{{{{\bf{g}}_{i,j}}}} - {q_t}), \notag
\end{align}
where $T$ denotes quantization bits and $q_t$ represents the quantized discrete value. Since we have used binarized quantization method, $T$ is set to 2 and $q_t \in \{ -1,1 \}$.  $f_\varepsilon$ denotes the density function of $\varepsilon$, where $\varepsilon = \sqrt {{p_i}} {{\bf{x}}_{i,j}} + \sum\limits_{c = 1,c \ne i}^N {\sqrt {{{p}_c}} {{\bf{x}}_{c,j}}}  + \frac{{{{\bf{n}}_{i,j}}}}{{{{\bf{g}}_{i,j}}}} - {q_t}$, which follows a distribution that combined with the known Bernoulli distribution and Cauchy distribution.

The training procedures of the SFDMA scheme is summarized in
Algorithm 1.

\begin{algorithm}[h]
	\caption{The proposed training algorithm for SFDMA BC network with inference tasks}
	\label{alg2}
	\KwIn{$T$ (number of epochs), $L$ (number of encoding representation samples per data sample)}
	\For{epoch $t=1$ \KwTo $T$}{
		Sample mini-batch of data samples $\left\{ {\left( {{{\bf{s}}_i},{{\bf{u}}_i}} \right)} \right\}_{i = 1}^N$ from each BS\;
		Compute conditional distribution  ${p_{{\phi _i}}}\left( {\left. {{{\bf{y}}_i}} \right|{{\bf{s}}_i}} \right)$\;
		Compute the conditional entropy term $H_{{\phi _i}}\left( {{Y_{i,j}}|{{\bf{s}}_i}} \right)$\;
		Quantize the discrete representation $\left\{ {z_j^{\left( {i,l} \right)}} \right\}_{l = 1}^L$ based on Eq.~\eqref{quantizer}\;
		Compute the entropy terms $\left\{ H\left( {{Y_{i,j}} \left| {{\bf{x}}_j^{\left( {i,l} \right)}} \right.} \right) \right\}_{l = 1}^L$\;
		Sample the noise $\left\{\boldsymbol{\epsilon}^{(i, l)}\right\}_{l=1}^L \sim \mathcal{C N}\left(0, \sigma^2 \mathbf{I}\right)$\;
		Estimate the corrupted discrete representation $\left\{ {{\bf{y}}_j^{\left( {i,l} \right)}} \right\}_{l = 1}^L$\;
		Compute the loss $\mathcal{L}_{V R I B}\left(\phi_i, \theta_i\right)$ based on Eq.~\eqref{VRIB2}\;
		Update the parameters $\phi_i$ and $\theta_i$ through backpropagation\;
	}
\end{algorithm}

  The training process not only minimizes semantic decoding error but also implicitly fosters the orthogonality of the encoded signals. On one hand, during joint training, the SFDMA naturally learns to semantic encoding   such  that the outputs are as different as possible to reduce redundancy. This drives the semantic encoded signals towards orthogonality, as orthogonal signals do not interfere with each other, allowing the decoders to separate them more easily. As training progresses, gradient descent optimization will push the semantic encoded signals towards directions that minimize mutual interference, eventually leading them towards an orthogonal solution. This is a spontaneous mechanism where, even without explicit orthogonality constraints, the training process reduces correlation between signals to minimize reconstruction error.

On the other hand, orthogonality reduces interference. Specifically, orthogonal signals have a zero inner product, meaning they do not overlap in signal space. By encouraging orthogonality, the system can minimize interference, improving decoder performance and making it easier to recover the original inputs.
The loss function  aims to minimize the semantic decoding error.   Given  the received   signals, if the semantic encoded signals interfere too much, it will result in poor reconstruction, increasing the loss. To minimize the loss, the training process will encourage the encoders to produce the semantic encoded signals that are well-separated in signal space, thus driving them toward orthogonality.

\section{ SFDMA for Image Reconstruction Semantic BC Network}

\begin{figure}[h]
	\centering
	\includegraphics[width=9cm]{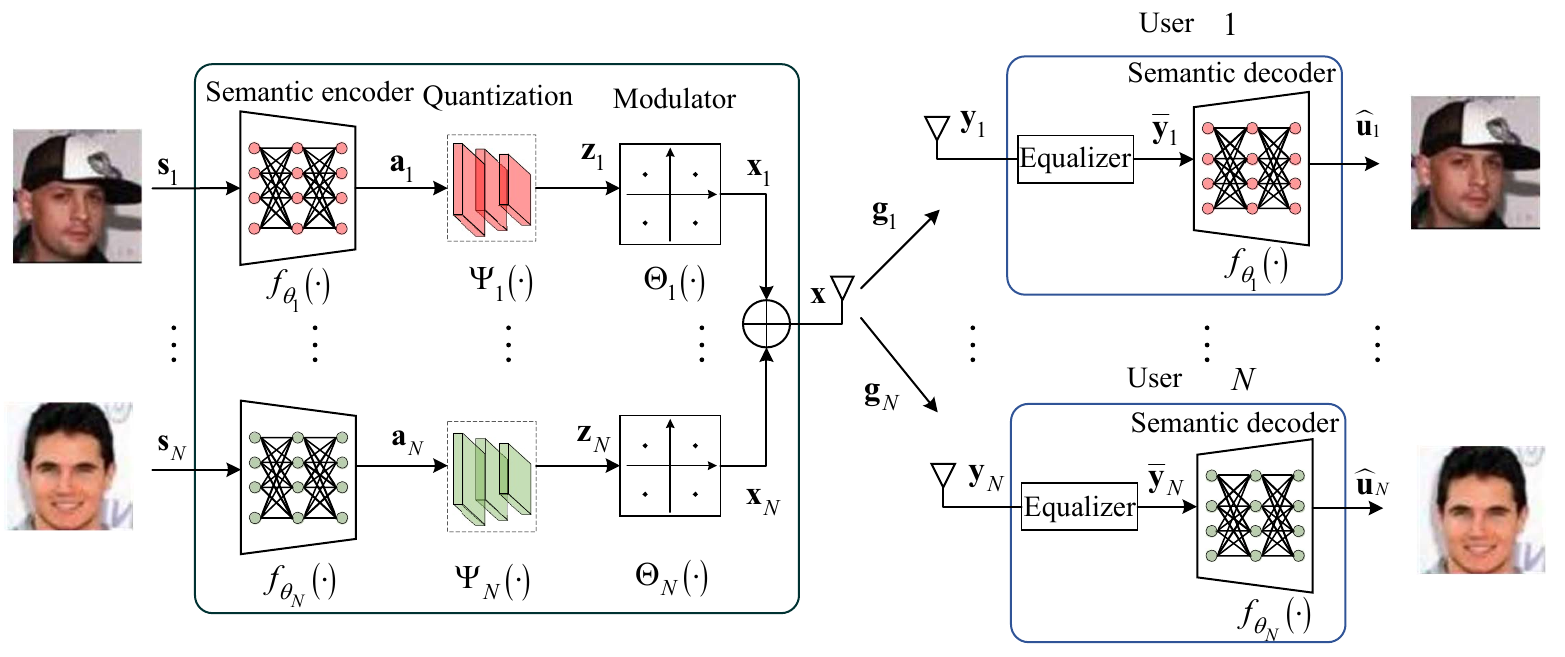}
	\caption{ A SFDMA BC network for image reconstruction }
	\label{system_model} 

\end{figure}

Furthermore, we investigate SFDMA scheme    for multi-user semantic BC networks with image reconstruction tasks.
Specifically, as shown in   Fig.~\ref{system_model}, by exploiting the SFDMA scheme, a semantic BS aims to transmit images   $\left\{ {{{\bf{s}}_i}} \right\}_{i = 1}^N$ to $N$ users simultaneously, where ${{\bf{s}}_i}$ is the intended image of  User $i$, $i=1,...,N$.
The semantic BS
includes semantic encoders $ {f_{{\varphi  _i}}}\left(  \cdot  \right)$, quantizers ${\Psi _i}\left(\cdot  \right)$ and modulators $\Theta _{\rm{m}}^i\left( {\cdot} \right)$.

\begin{figure}[htb]
	\centering
	\begin{subfigure}[ SFDMA encoder ${f_{{\varphi _i}}}$ for User  $i$.]{%
			\includegraphics[width=7cm]{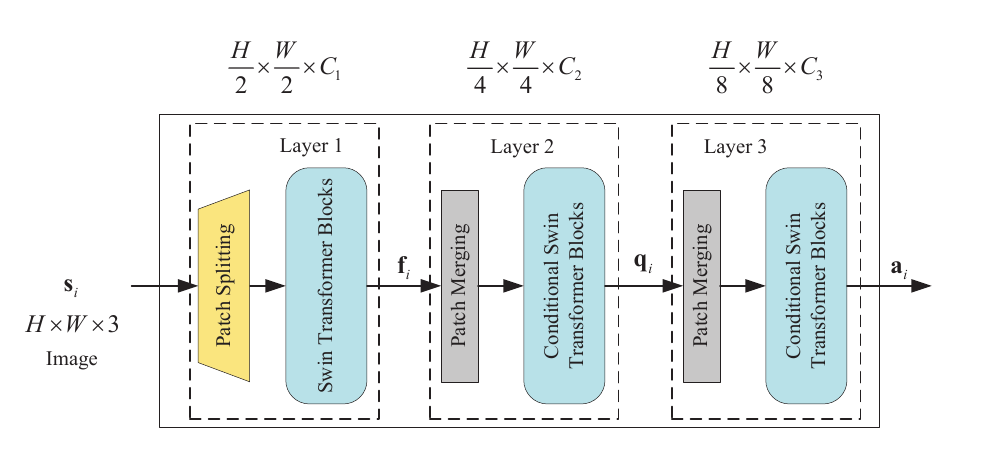}
			\label{feature extraction}}
	\end{subfigure}
	\qquad
	\qquad
	\qquad
	\qquad
	\qquad
	\begin{subfigure}[ Semantic decoder ${f_{{\theta _i}}}$ of User $i$. ]{%
			\includegraphics[width=7cm]{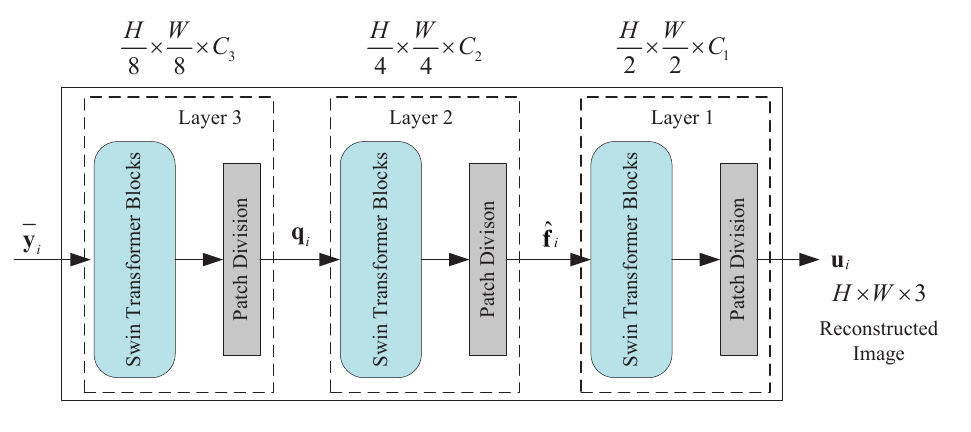}
			\label{semantic recovery}}
	\end{subfigure}
	\caption{(a) The   architecture of the semantic encoder ${f_{{\phi _i}}}$  of SFDMA  BC network for image reconstruction; (b)  The   architecture of the semantic decoder of User $i$ for image reconstruction}
	\label{fig3}
\end{figure}

Fig.~\ref{feature extraction} shows the architecture of the semantic encoder of   ${f_{{\varphi _i}}}\left(  \cdot  \right)$ for User $i$  with the parameter set $ {\varphi _i} $.
The semantic encoder  ${f_{{\varphi _i}}}\left(  \cdot  \right)$
includes three layers, where the dimensions are 128, 256 and 512, respectively.

In Layer 1, an RGB image ${{\bf{s}}_i} \in {{\mathbb R}^{3 \times H \times W}} $ is first split into ${l_1} = \frac{H}{2} \times \frac{W}{2}$  non-overlapping patches. After patch embedding, Swin Transformer blocks are applied to these $l_1$ non-overlapping patches. Here, we refer to these $N_1$ blocks together with the patch embedding layer as
\begin{align}\label{1}
{{\bf{f}}}_i = {\rm{Layer1}}\left( {{{\bf{s}}_i}} \right),
\end{align}
where ${{\bf{f}}_{\bf{i}}} \in {{\mathbb R}^{\frac{H}{2} \times \frac{W}{2} \times {C_1}}}$ is fed to Layer 2 for downs-sampling through a patch merging layer, and the down-sampling data is processed by Swin Transformer blocks, i.e.,
\begin{align}\label{2}
{{\bf{q}}_i} = {\rm{Layer2}}\left( {{{\bf{f}}_i}} \right),
\end{align}
where ${{\bf{q}}_i} \in {{\mathbb R}^{\frac{H}{4} \times \frac{W}{4} \times {C_2}}}$ denotes the output of Layer 2. Furthermore, $\left\{ {\mathbf{q}_i} \right\} _{i=1}^N$ are processed by Layer 3, which includes a down-sampling Patch Merging layer and Swin Transformer blocks. Finally, the semantic encoded feature vector $\mathbf{a}_i \in \mathbb{R}^{\frac{H}{8} \times \frac{W}{8} \times C_3}$ is given by
\begin{align}\label{3}
{{\bf{a}}_i} = {\rm{Layer3}}\left( {{{\bf{q}}_i}} \right).
\end{align}

In summary, the semantic encoder of SFDMA scheme can be formulated as
\begin{align}\label{img1}
{{\bf{a}}_i} = {f_{{\varphi _i}}}\left( {\bf{s}}_i \right),
\end{align}
where ${\bf{a}}_i$ is the encoded semantic features vectors from ${\bf{s}}_i$, and $ {f_{{\varphi _i}}}\left( \cdot \right) $ is the semantic encoder with the parameter set $ {\varphi _i} $.

 Then, the analog semantic features $\left\{ {\bf{a}}_i \right\}_{i=1}^N$ are quantized as
\begin{align}
\mathbf{z}_i={\Psi _i}\left( {{{\bf{a}}_i}} \right),~i\in\{1,..,N\},
\end{align}
where ${{\bf{z}}_i}$ is the quantized signal. Then, we adopt BPSK modulation
\begin{align}\label{000}
{{\bf{t}}_i} = \Theta _{{i}}\left( {{{\bf{z}}_i}} \right),~i\in\{1,..,N\}.
\end{align}

Before transmission, we perform normalization processing,
\begin{align}\label{2}
{{\bf{x}}_i} = \frac{{{{\bf{t}}_i}}}{{{{\left\| {{{\bf{t}}_i}} \right\|}_2}}}.
\end{align}

For User $i$,  the received signal is computed based on Eq.~\eqref{00}, and then we use Eq.~\eqref{received_signal} to obtain the equalized signal ${\overline {\bf{y}} _i}$.

Furthermore, the equalized signal ${\overline{\bf{y}}_i}$ is sent to semantic decoder ${f_{{\theta _i}}}\left( \cdot \right)$, as shown in  Fig.~\ref{semantic recovery}, which  consists of three layers.
 Specifically, in Layer $1$, the feature vector $\mathbf{{ {\bf{y}}}}_i \in \mathbb{R}^{\frac{H}{8} \times \frac{W}{8} \times C_3}$ is first sampled by Swin Transformer blocks. Then by applying a Patch Division layer, we obtain $\hat{\mathbf{q}}_i$ as
\begin{align}\label{000}
{\widehat {\bf{q}}_i} = {\rm{Layer3}}\left( {{{\overline {\bf{y}} }_i}} \right).
\end{align}

Then, $\hat{\mathbf{q}}_i \in \mathbb{R}^{\frac{H}{4} \times \frac{W}{4} \times C_2}$ is fed to Layer 2, which includes Swin Transformer blocks and a up-sampling Patch Division layer. The output of Layer 2 is given by
\begin{align}\label{000}
{\widehat {\bf{f}}_i} = {\rm{Layer2}}\left( {{{\widehat {\bf{q}}}_i}} \right){\rm{,}}
\end{align}
where  $\hat{\mathbf{f}}_i \in \mathbb{R}^{\frac{H}{2} \times \frac{W}{2} \times C_1}$ is sent into Swin Transformer blocks and an up-sampling Patch Division layer. The  reconstructed  image      $\hat{\mathrm{u}}_i \in \mathbb{R}^{H \times W \times 3}$ is given as
\begin{align}\label{000}
{\widehat {\bf{s}}_i} = {\rm{Layer1}}\left( {{{\widehat {\bf{f}}}_i}} \right).
\end{align}

Therefore, the semantic decoder of User $i$ is given as
\begin{align}\label{img4}
{\widehat {\bf{s}}_i} = {f_{{\theta _i}}}\left( {{{\widehat {\bf{z}}}_i}} \right),i \in \{ 1,..,N\} ,
\end{align}
where ${f_{\theta _i} }\left( {\cdot} \right)$ denotes the semantic decoder of User $i$ with its
parameters $\theta_i $.

Moreover,  we adopt the MSE loss function for image reconstruction of the proposed SFDMA scheme as follows
\begin{align}\label{mse}
{\mathcal{L}\left( {{{\bf{s}}_i},{{\widehat {\bf{s}}}_{i}}} \right) = \frac{1}{n}\sum\limits_{j = 1}^n {{{\left( {{{\bf{s}}_{i,j}} - {{\widehat {\bf{s}}}_{i,j}}} \right)}^2}} ,}
\end{align}
where $n$ denotes the number of images.
In summary,
the training procedure for SFDMA BC   network with image reconstruction tasks is summarized in Algorithm~\ref{alg22}.
\begin{algorithm}[h]
	\caption{Training algorithm for SFDMA BC network with image reconstruction tasks}
	\label{alg22}
	\KwIn{$T$ (number of epochs), $N$ (number of users), noise ${\bf{n}}_i$ with a fixed SNR value}
	\For{epoch $t=1$ \KwTo $T$}{
		\textbf{Base station}\;
		\Indp
		Sample mini-batch of data samples $\left\{ {\left( {{{\bf{s}}_i},{{\bf{u}}_i}} \right)} \right\}_{i = 1}^N$ from each BS\;
		Compute the feature vectors $\left\{ {f_{{\phi _i}}\left( {{{\bf{s}}_i}} \right)} \right\}_{i = 1}^N \to \left\{ {{{\bf{a}}_i}} \right\}_{i = 1}^N$\;
		Compute the quantized vectors $\left\{ {\Psi _i\left( {{{\bf{a}}_i}} \right)} \right\}_{i = 1}^N \to \left\{ {{{\bf{z}}_i}} \right\}_{i = 1}^N$\;
		Compute the modulation signal $\left\{ \Theta _{{i}}\left( {{{\bf{z}}_i}} \right) \right\}_{i = 1}^N \to \left\{ {{{\bf{x}}_i}} \right\}_{i = 1}^N$\;
		\Indm
		
		\textbf{Semantic User $i$}\;
		\Indp
		Compute the equalization received signal $\left\{ {{{\bf{y}}_i}} \right\}_{_{i = 1}}^N \to \left\{ {\overline{{{\bf{y}}_i}}} \right\}_{i = 1}^N$\;
		Image reconstruction $\left\{ {f_{{\theta _i}}\left( {\overline{{{\bf{y}}_i}}} \right)} \right\}_{i = 1}^N \to \left\{ {{{\widehat {\bf{s}}}_i}} \right\}_{i = 1}^N$\;
	    Compute the loss MSE based on Eq.~\eqref{mse}\;
		Update the parameters $\phi_i$ and $\theta_i$ through backpropagation\;
		\Indm
	}
\end{algorithm}

\section{Optimal Power Allocation Scheme for SFDMA BC Network}

The performance of semantic BC network depends on the transmit  power. However,  the   relationship between the performance of semantic communications and the transmit  power has not yet been established,  which leads to no theoretical basis for the power allocation for  the  semantic BC network.
To overcome this   challenge, we establish the  relationship between performance and SINR for the  SFDMA BC network.

Let
 ${\phi _i}$  and   ${\rm{SIN}}{{\rm{R}}_i}$ denote the  performance   and received   SINR   of User $i$, respectively. Note that the performance is related to the task at hand, e.g. in image classification tasks the performance metric is accuracy, in image reconstruction tasks the performance is peak signal-to-noise ratio (PSNR) or multiscale
 	structural similarity index metric (MS-SSIM), etc.

\begin{figure}[htp]
	\centering
	\begin{subfigure}[ ABG formula  for  inference task. ]{%
			\includegraphics[width=5.5cm]{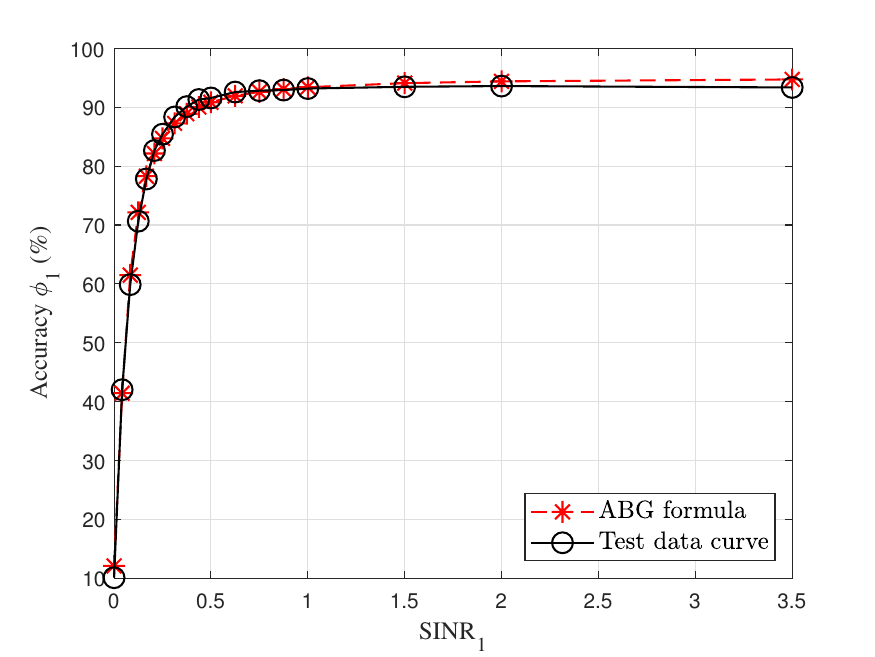}
			\label{ABG_1}}
	\end{subfigure}
	\begin{subfigure}[ ABG formula for   image reconstruction task. ]{%
			\includegraphics[width=5.5cm]{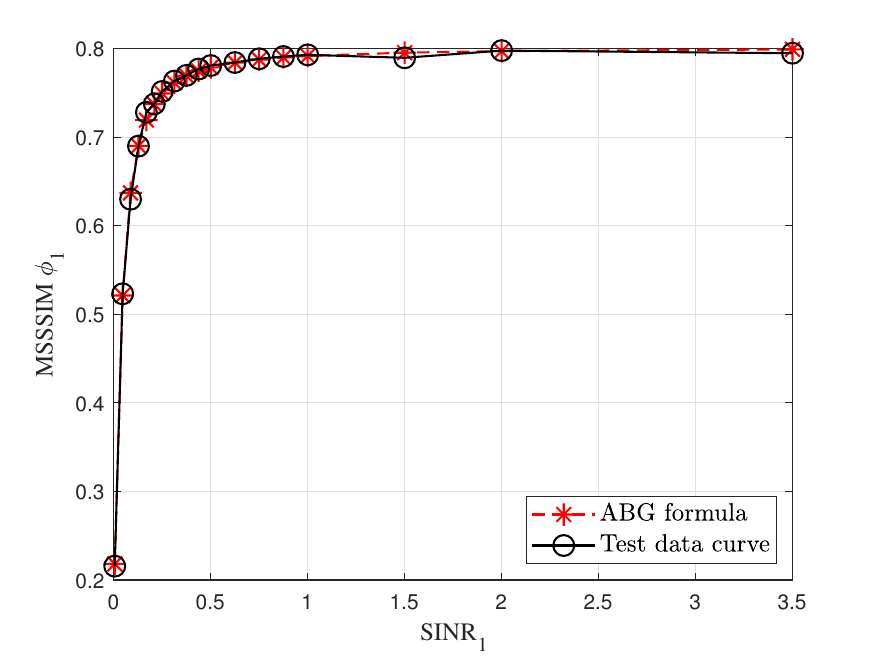}
			\label{ABG_2}}
	\end{subfigure}
	\caption{ Comparison between the   ABG formula and the test data curves. }
	\label{ABG}
\end{figure}

With extensive experiments, we find that the performance and   SINR satisfies a specific law: \emph{as SINR increases, performance  increases rapidly at first, and then slowly increases until it reaches the upper bound}, for example, in the image  classification task on the MNIST dataset, the relationship between classification accuracy and SINR is shown in Fig.~\ref{ABG_1}. Additionally, the relationship between MS-SSIM and SINR is also illustrated in Fig.~\ref{ABG_2}.


 \begin{table}[t]
	\caption{Fitting parameters.}
	\centering
	\scalebox{0.8}{
		\begin{tabular}{|c|c|c|c|c|c|}
			\hline
			\rule{0pt}{15pt}Parameters of ABG	formula &${\alpha _1}$ &${{\beta _1}}$ & ${{\gamma _1}}$ & ${{\tau _1}}$&$\zeta $ \\ \hline
			\rule{0pt}{15pt}  Values & 95
			&15.7 &82.93 & 1.427 &0.6338\\ \hline
		\end{tabular}
		\label{Parameter_ABG}
	}
\end{table}


Therefore, the  relationship between ${\phi _i}$    and  ${\rm{SIN}}{{\rm{R}}_i}$  of User $i$     can be  fitted by ABG function (as shown in Fig.~\ref{ABG}) and is given by
\begin{subequations}
\begin{align}
 &{\phi _i}\left( {\left\{ {{p_i}} \right\}_{i = 1}^N} \right){\rm{ = }}{\alpha _i} - \frac{{{\gamma _i}}}{{1 + {{\left( {{\beta _i}{\rm{SIN}}{{\rm{R}}_i}} \right)}^{{\tau _i}}}}},
\\
&{\rm{SIN}}{{\rm{R}}_i} = \frac{{{p_i}{{\left| {{\mathbf{g}_i}} \right|}^2}}}{{\sum\limits_{j = 1,j \ne i}^N {{p_j}{{\left| {{\mathbf{g}_i}} \right|}^2}}  + \sigma _i^2}},
\end{align}
\end{subequations}
where ${\alpha _i}$, ${{\beta _i}}$, ${{\gamma _i}}$ and ${{\tau _i}}$ are parameters of the ABG formula.
Given the DL based semantic encoders and decoders, the
parameters $\left\{ {{\alpha _i},{\beta _i},{\gamma _i},{\tau _i}} \right\}_{i = 1}^N$ can be obtained through testing, as listed  in Table 	\ref{Parameter_ABG}.

 With the ABG function, we  optimize power allocation of the  SFDMA BC network to minimize the total power consumption under the  performance constraints of the $N$ users. Mathematically,  the  power allocation  optimization  problem of the  SFDMA BC network can be  formulated as

  \begin{subequations}
	\begin{align}\label{optimize}
		\mathop {\min }\limits_{\left\{ {{p_i}} \right\}_{i = 1}^N} &\sum\limits_{i = 1}^N {{p_i}} \\
		{\rm{s}}{\rm{.t.}}&{\phi _i}\left( {\left\{ {{p_i}} \right\}_{i = 1}^N} \right) \ge {\eta _i},i = 1,...,N,\label{ABG_b}
	\end{align}
\end{subequations}
where ${\eta _i}$ denotes the performance requirement of User $i$.

Furthermore, constraint \eqref{ABG_b} can be equivalently reformulated as linear constraint as follows
   \begin{align}{p_i}{\left| {{g_i}} \right|^2} \ge \frac{1}{{{\beta _i}}}{\left( {\frac{{{\gamma _i}}}{{{\alpha _i}\left( {{n_{\rm{b}}}} \right) - {\eta _i}}} - 1} \right)^{\frac{1}{{{\tau _i}}}}}\left( {\sum\limits_{j = 1,j \ne i}^N {{p_j}{{\left| {{g_i}} \right|}^2}}  + \sigma _i^2} \right)\label{constraint}\end{align}

Thus, the    power allocation  optimization  problem of the  SFDMA BC network  is a linear programming, which can be efficiently solved by simplex method\cite{li2007distributed}.

Therefore, the workflow of the task-oriented  SFDMA BC network is summarized in Algorithm~\ref{workflow}.

\begin{algorithm}[h]
	\caption{ Task-oriented SFDMA BC network workflow }
	\label{workflow}
	\KwIn{ $\left\{ {\left( {{{\bf{s}}_i},{{\bf{u}}_i}} \right)} \right\}_{i = 1}^N$, $N$ (number of users), $\left\{\sigma _i^2\right\}_{i = 1}^N$ (noise variance), $\left\{\mathbf{g} _i^2\right\}_{i = 1}^N$ (channel gain), $\left\{ \psi_i \right\}_{i = 1}^N$ (parameters of encoders), $\left\{ \theta_i \right\}_{i = 1}^N$ (parameters of decoders)}
	\KwOut{$\left\{ {{{\bf{\hat{s}}}_i}} \right\}_{i = 1}^N$ or $\left\{ {{{\bf{\hat{u}}}_i}} \right\}_{i = 1}^N$}
	
	Obtain the semantic feature $\left\{ \mathbf{x}_i \right\}_{i = 1}^N$ from the raw data $\left\{ \mathbf{s}_i \right\}_{i = 1}^N$ using the encoder\;
	Compute the allocated power $\left\{ {{{p}_i}} \right\}_{i = 1}^N$ according to Eq.~\eqref{optimize} and \eqref{constraint}\;
	Transmit the semantic information of multiple users, the received signal of user $i$ is given as Eq.~\eqref{received_signal}\;
	Equalize the received data according to Eq.~\eqref{eq_received_signal}\;
	\eIf{the current task is an inference task}
	{Infer the semantic information $\left\{ {{{\mathbf{\hat{u}}}_i}} \right\}_{i = 1}^N$ through the decoder}
	{Reconstruct the original data $\left\{ \mathbf{\hat{s}}_i \right\}_{i = 1}^N$ through the decoder}
\end{algorithm}

\section{Experimental Results and Analysis}
\subsection{ Experimental Setup}

In this section, we perform a comprehensive analysis
through the comparison between the proposed SFDMA BC schemes and
the conventional deep JSCC schemes\cite{xie2021deep}. While both use similar neural network designs, the SFDMA BC scheme uniquely incorporates SFDMA, effectively mitigating multi-user interference.
The key experimental parameters  are shown in Table~\ref{settings}.

\begin{table}[ht]
	\caption{Key parameters   of the SFDMA BC network.}
	\centering
	\begin{tabular}{|c|c|c|}
		\hline
		Tasks & Reconstruction & Classification \\ \hline
		Batch size & 32 & 512 \\ \hline
		Latent dimensions & 512 & 1024 \\ \hline
		Learning rate & 1e-4 & 1e-4 \\ \hline
		Loss function & MSE & RIB \\ \hline
		Training SNR & 0dB & 0dB \\ \hline
		Test metrics & MS-SSIM & Accuracy \\ \hline
		Threshold & 0.77 & 92 \\ \hline
	\end{tabular}
	\label{settings}
\end{table}

In    inference tasks  experiment, the semantic encoder consists of a dense layer and  a   tanh activation function, and  the semantic decoders consist  of three dense layers, two ReLU as activation function and a softmax function. In the  inference tasks networks, the learning rate is $\delta  = {10^{ - 3}}$, and the weights of models are updated by the Adam
optimizer.
The inference tasks experiments are conducted using the widely-used
MNIST dataset\cite{deng2012mnist}, which comprises a training set of 60,000 sample images and a test
set of 10,000 sample images across 10 handwritten digits categories.
Meanwhile, the CelebA dataset\cite{liu2015faceattributes}, which includes 202,599 face images and 10,177 celebrity identities, is used for the image reconstruction
tasks experiments.

We compare the proposed SFDMA BC scheme with conventional deep JSCC scheme, where the multi-user interference is not ignored during the   training phase but is considered during the test phase.
Moreover,   the   deep JSCC scheme without multi-user interference in both  training and testing phases are
also used for comparison, which is named Upper Bound.

\subsection{Performance evaluation for inference tasks}
\begin{figure}[htp]
	\centering
	\quad
	\begin{subfigure}[Two users ($N=2$).]{%
			\includegraphics[width=5.0cm]{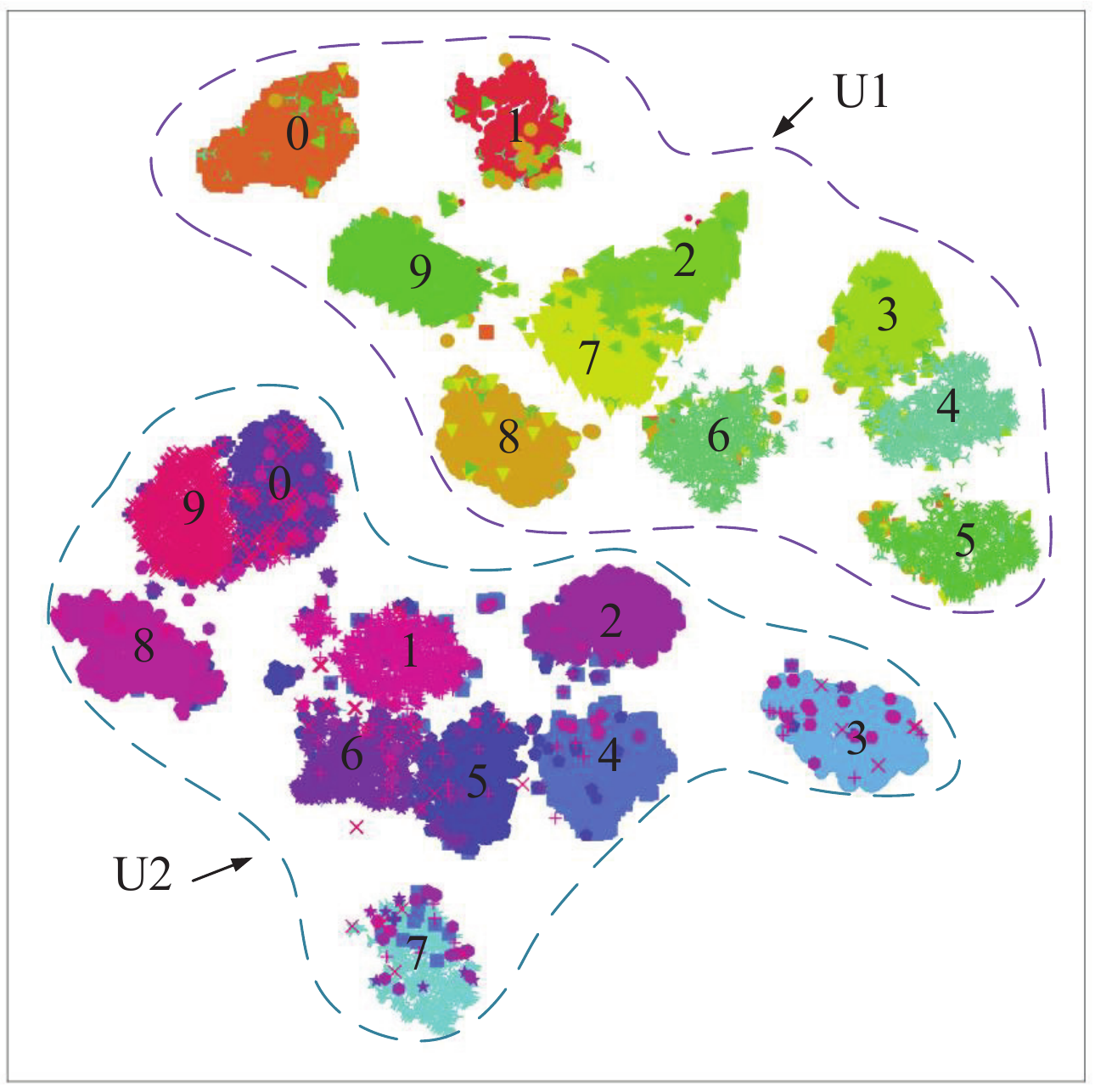}
			\label{class2}}
	\end{subfigure}
	\begin{subfigure}[Three users ($N=3$).]{%
			\includegraphics[width=5.0cm]{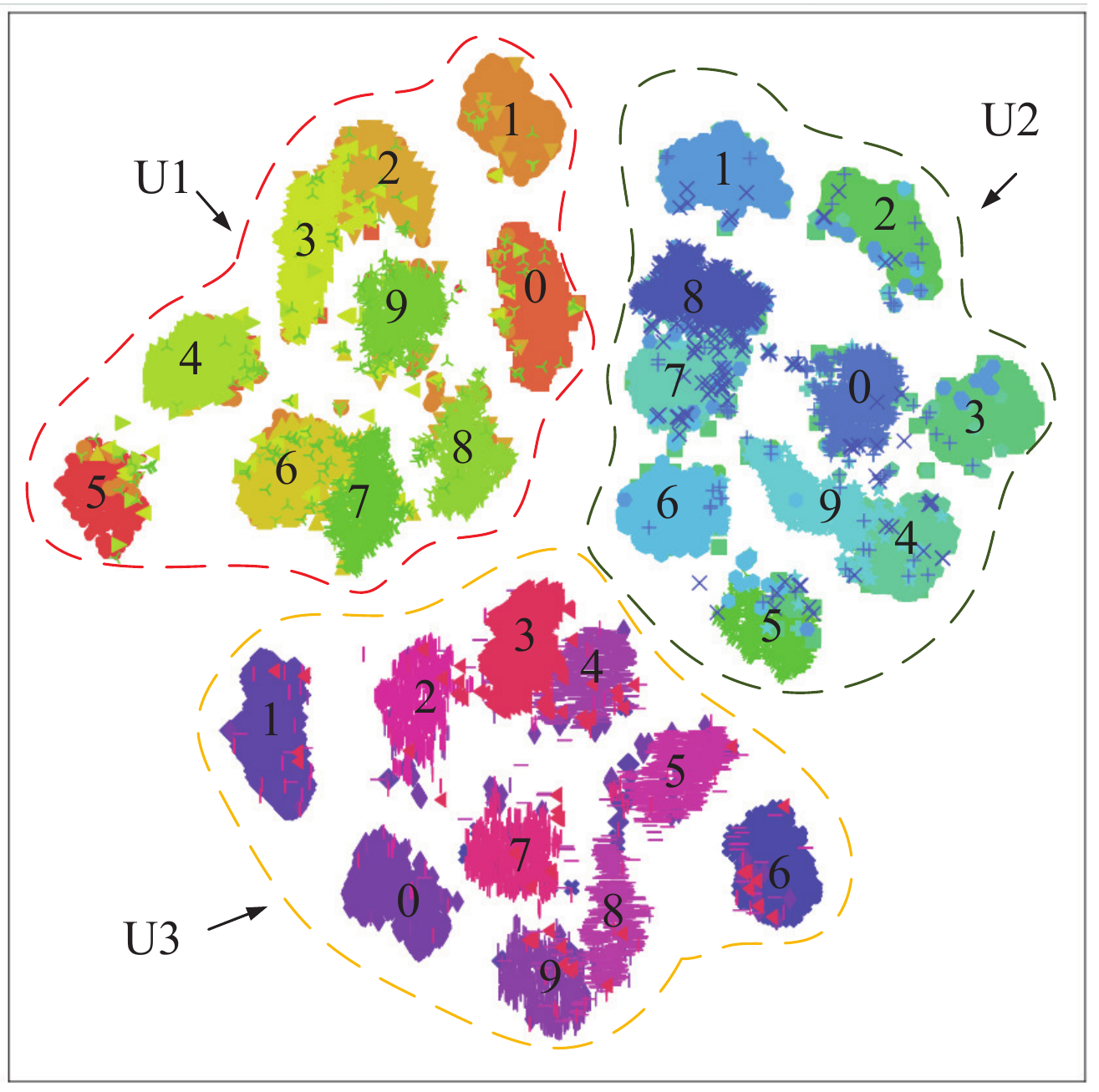}
			\label{class3}}
	\end{subfigure}
	\caption{ Two-dimensional t-SNE embedding of the  semantic  features  $\left\{ {{{\bf{x}}_i}} \right\}_{i = 1}^N$ of the proposed SFDMA scheme with  ${q_{{\rm{bit}}}} = 128$ bits and  SNR = 5dB }
	\label{fig4}
\end{figure}

   First, we evaluate the
semantic features separation performance  of the
proposed SFMDA BC scheme by adopting the dimension reduction technique
of t-SNE (t-Distributed Stochastic Neighbor Embedding) \cite{van2008visualizing}, which is a powerful dimensionality reduction technique that excels in preserving the local structure of high-dimensional data when projecting it into a lower-dimensional space. By visualizing the semantic encoded feature vectors in 2-dimensional space using t-SNE, we can observe how well-separated the vectors are, which reflects their orthogonality in the original high-dimensional space.

\begin{figure}[htb]
	\centering
	\begin{subfigure}[Quantization bits ${q_{{\rm{bit}}}} = 32$ bits]{%
			\includegraphics[width=6.5cm]{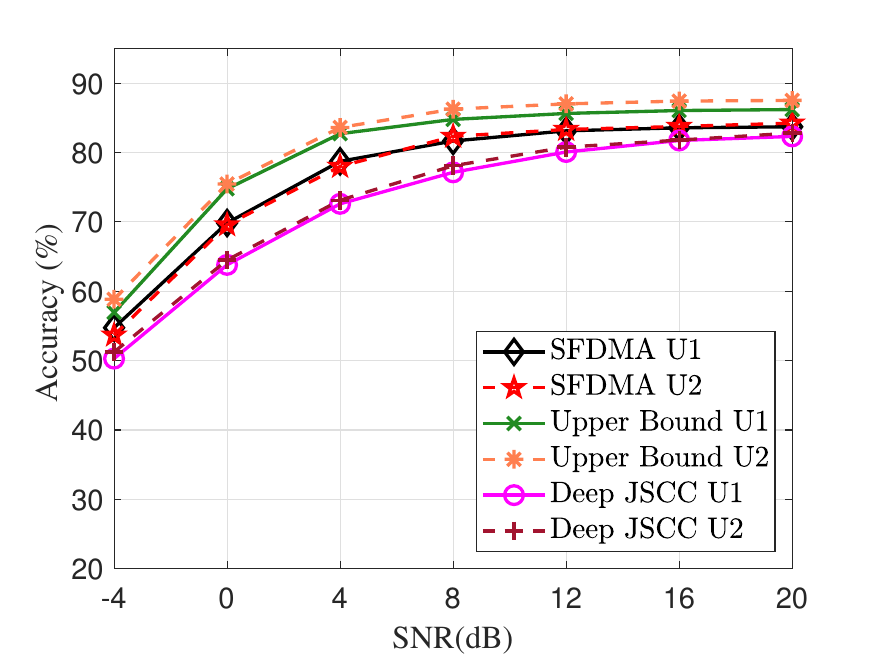}
			\label{taskb1}}
	\end{subfigure}
	\begin{subfigure}[Quantization bits ${q_{{\rm{bit}}}} = 64$ bits]{%
			\includegraphics[width=6.5cm]{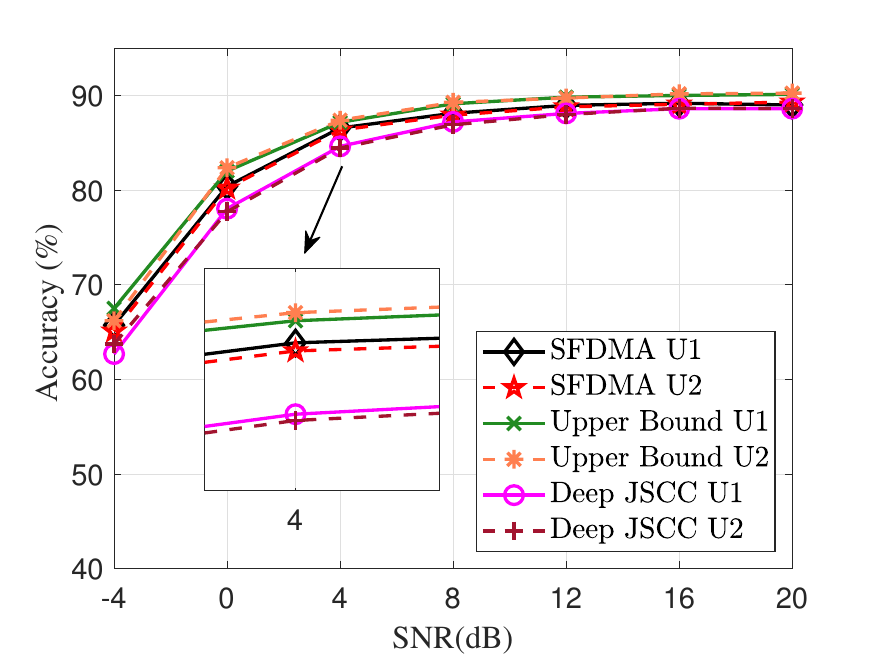}
			\label{taskb2}}
	\end{subfigure}
	\caption{Performance of classification accuracies  over Rayleigh channels with training SNR=0dB.
	}
	\label{task21}
\end{figure}

 Fig.~\ref{class2} and   \ref{class3}  illustrate two-dimensional t-SNE embedding of
  the  semantic features     of the proposed SFMDA BC schemes with two users and three users respectively,   where the  quantization bits ${q_{{\rm{bit}}}} = 128$ bits and the training SNR = 5dB.
For the two users SFDMA BC network, the  classification accuracy of User $1$ and $2$ are $93.1\%$ and $93.4\%$, respectively, and   for the two users scenario, the  classification accuracy of User $1$, $2$ and $3$  are $93.23\%$, $94.14\%$ and $93.4\%$, respectively, which demonstrate the effectiveness of the proposed SFDMA scheme. In
 Fig.~\ref{class2} and  \ref{class3}, each point corresponds to the position of a semantic feature in 2-dimensional space obtained through t-SNE, with different colors representing different classes and users. The semantic features belonging to the same user are highlighted by dashed enclosures, the absence of overlap between the semantic features of different users show  that the semantic features $\left\{ {{{\bf{x}}_i}} \right\}_{i = 1}^N$  are approximately separated, which verifies the approximate orthogonality property of semantic features of the proposed SFDMA scheme.

\begin{table}[tp]
	\caption{ The inner product and angle among the semantic features $\left\{ {{{\bf{x}}_i}} \right\}_{i = 1}^N$  with the same inputs.}
	\centering
	\scalebox{0.8}{
		\begin{tabular}{|c|c|c|c|}
			\hline
			\rule{0pt}{15pt} Inner product& ${\mathbf{x}}_1^H{{\mathbf{x}}_2}$& ${\mathbf{x}}_1^H{{\mathbf{x}}_3}$ & ${\mathbf{x}}_2^H{{\mathbf{x}}_3}$ \\ \hline
			
			\rule{0pt}{15pt}  Value&$1.0  \times {10^{ - 3}}$ &$2.5 \times {10^{ - 3}}$&$ 2.6 \times {10^{ - 4}}$ \\ \hline
			\rule{0pt}{15pt}   Angle& $\arccos \left( {\mathbf{x}}_1^H{{\mathbf{x}}_2} \right)$& $\arccos \left( {\mathbf{x}}_1^H{{\mathbf{x}}_3} \right)$ & $\arccos \left( {\mathbf{x}}_2^H{{\mathbf{x}}_3} \right)$\\ \hline
			
			\rule{0pt}{15pt}  Value$\left( {^ \circ } \right)$&$89.94^ \circ$ &$89.86^ \circ$&$ 89.99^ \circ$ \\ \hline	
		\end{tabular}
		\label{table2}
	}
\end{table}

Table~\ref{table2} illustrates the inner product and the angle among the semantic features ${\bf{x}}_1$, ${\bf{x}}_2$ and ${\bf{x}}_3$.
Table~\ref{table2} shows that even with the same inputs, the inner products among the semantic features of different users are $1.0  \times {10^{ - 3}}$, $2.5 \times {10^{ - 3}}$, and $ 2.6 \times {10^{ - 4}}$, which approach zero. Moreover, the corresponding angles are $89.94^ \circ$, $89.86^ \circ$ and $ 89.99^ \circ$, which approach ${\rm{90}}^ \circ $.
  The results in Table~\ref{table2} verify    the semantic features of multi-user
are approximately orthogonal  for the proposed SFDMA scheme.

\begin{table}[tp]
	\caption{The     classification accuracy of the SFDMA  BC  network with the same inputs.}
	\centering
	\scalebox{0.8}{
	\begin{tabular}{|c|c|c|c|c|}
		\hline
		\rule{0pt}{15pt}	&$f_{\theta_{1}}(\mathbf{y}_{1})$ &$f_{\theta_1}\left(\mathbf{g}_{1},\mathbf{x}_2\right)$ & $f_{\theta_{2}}(\mathbf{y}_{2})$ & $f_{\theta_{2}}\left(\mathbf{g}_{2},\mathbf{x}_{1}\right)$ \\ \hline
		\rule{0pt}{15pt}  	Accuracy(\%) & 91.91 &9.66& 91.89  &9.71 \\ \hline
	\end{tabular}
	\label{table1}
}
\end{table}

Table~\ref{table1} illustrates the classification accuracy   of  two users with  the same inputs.
 As shown in  Table~\ref{table1},  even with the same inputs, each user can accurately decode the intended semantic information, and the classification accuracy of User 1 and 2 are  $91.91\%$ and $91.89\%$, respectively. Meanwhile, each user cannot effectively decode the semantic information of other users, i.e.,     the classification accuracy  of  User $1$ decoding User 2's semantic information is only $9.66\%$,  the classification accuracy  of  User $1$ decoding User 2's semantic information is only $9.71\%$.
  The results in Table~\ref{table1} also verify    the semantic features of multi-user
are approximately orthogonal  for the proposed SFDMA scheme,
  Moreover, Table~\ref{table1} also demonstrates
 the proposed SFDMA can provide semantic confidentiality, and protect the  semantic information from being decoded by other users.

\begin{figure}[htb]
	\centering
			\includegraphics[width=6.5cm]{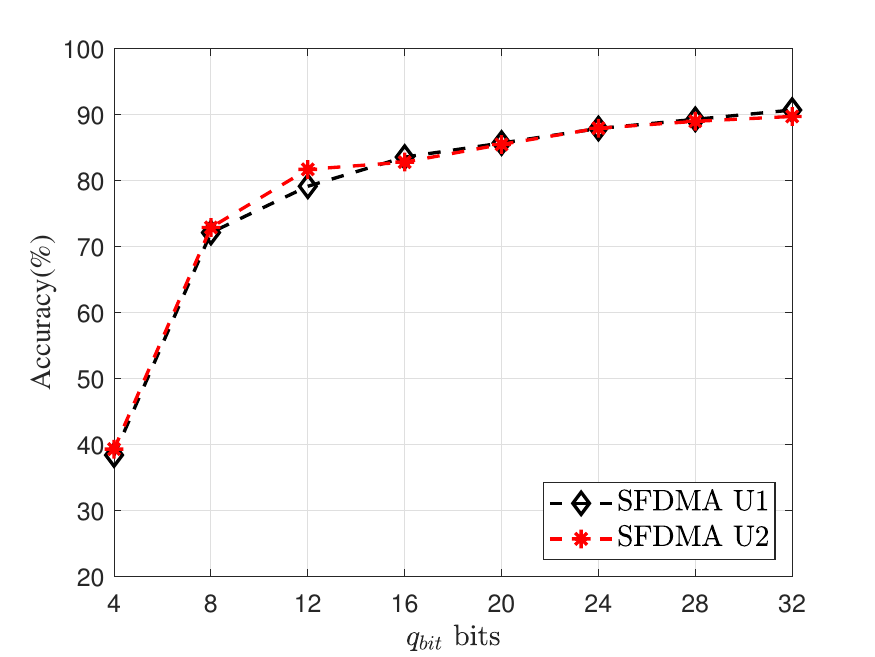}
			\label{fig:subfigure2}
		\caption{Performance of classification accuracies  versus  quantization bit    ${q_{{\rm{bit}}}}$    with training SNR=10dB.}
		\label{fig33}
	\end{figure}

Fig.~\ref{taskb1} and \ref{taskb2}
illustrate   classification accuracies of  Deep JSCC,  Upper Bound and the proposed SFDMA versus  SNR over  Rayleigh channel with quantization bits ${q_{{\rm{bit}}}} = 32$ and $64$ bits, respectively,  where the training SNR is 0dB.
Fig.~\ref{taskb1} and \ref{taskb2} show  that, as SNR
increases, the  classification accuracies of three schemes   increase  rapidly at first, and then slowly
increases until it reaches the maximum value.
Moreover, the classification accuracies of the proposed SFDMA scheme are higher than those of the Deep JSCC and are closer to the Upper Bound in comparison, which verify  the proposed SFDMA scheme can effectively eliminate multi-user interference  and improve inference tasks performance.
Comparing Fig.~\ref{taskb1} and \ref{taskb2}, the classification accuracies with  quantization bits ${q_{{\rm{bit}}}} = 64$ bits are higher than that with ${q_{{\rm{bit}}}} = 32$ bits. Thus,
a larger number of quantization bits ${q_{{\rm{bit}}}}$ leads to higher classification accuracy.

\begin{figure}[hbt]
	\centering
	\begin{subfigure}
		[PSNR1 = 21, PSNR2 = 21.45, MS-SSIM = 0.802, MS-SSIM = 0.806.]{%
			\includegraphics[width=5.2cm]{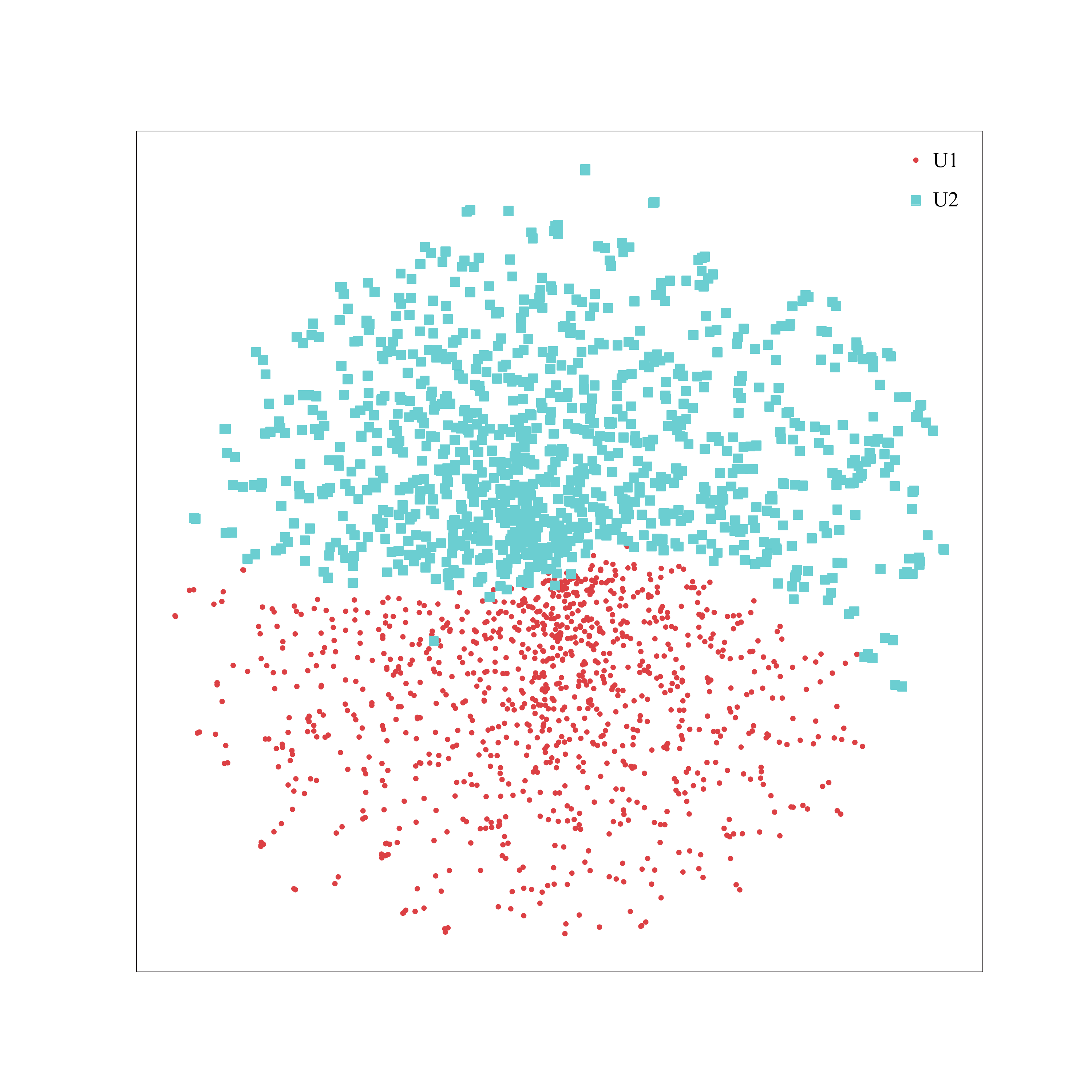}
			\label{fig:re1}}
	\end{subfigure}
\begin{subfigure}[PSNR1 = 24.76dB, PSNR2 = 24.93dB, PSNR3 = 25dB, MS-SSIM1 = 0.907, MS-SSIM2 = 0.909, MS-SSIM3 = 0.91.]{%
		\includegraphics[width=6.7cm]{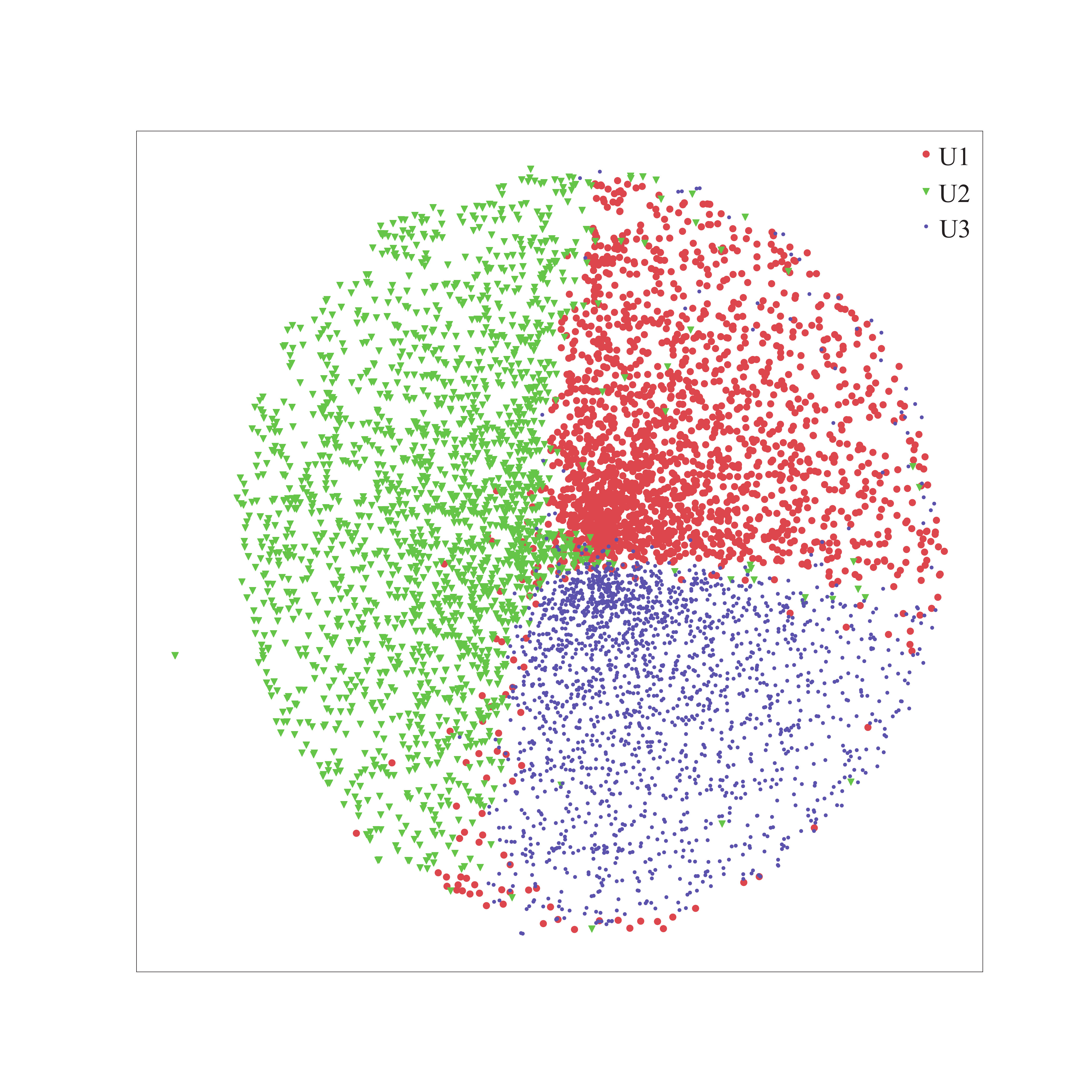}
		\label{fig:re3}}
\end{subfigure}
\caption {Two-dimensional t-SNE embedding of the received feature in the CelebA reconstruction task.}
\label{fig6}
\end{figure}

  Fig.~\ref{fig33} shows the classification accuracy versus  quantization bits ${q_{{\rm{bit}}}}$ with training SNR = $10 $dB. It can be seen from     Fig.~\ref{fig33} that
 as quantization bits ${q_{{\rm{bit}}}}$
increases, the  classification accuracy       increases  rapidly at first, and then slowly
increases until it reaches the upper bound.

\begin{table}[tb]
	\caption{ The inner product and angle among the semantic features of three transmission pairs  with the same inputs.}
	\centering
	\scalebox{0.8}{
		\begin{tabular}{|c|c|c|c|}
			\hline
			\rule{0pt}{15pt} Inner product& ${\mathbf{x}}_1^H{{\mathbf{x}}_2}$& ${\mathbf{x}}_1^H{{\mathbf{x}}_3}$ & ${\mathbf{x}}_2^H{{\mathbf{x}}_3}$ \\ \hline
			
			\rule{0pt}{15pt}  Value&$-1.0  \times {10^{ - 3}}$ &$7.0 \times {10^{ - 3}}$&$ -2.8 \times {10^{ - 4}}$ \\ \hline
			\rule{0pt}{15pt}   Angle& $\arccos \left( {\mathbf{x}}_1^H{{\mathbf{x}}_2} \right)$& $\arccos \left( {\mathbf{x}}_1^H{{\mathbf{x}}_3} \right)$ & $\arccos \left( {\mathbf{x}}_2^H{{\mathbf{x}}_3} \right)$\\ \hline
			
			\rule{0pt}{15pt}  Value$\left( {^ \circ } \right)$&$90.06$ &$89.60$&$ 90.02$ \\ \hline	
		\end{tabular}
		\label{tr}
	}
\end{table}

\begin{table}[t]
	\caption{The image reconstruction of the SFDMA BC network with the same input images.}
	\centering
	\scalebox{0.8}{
		\begin{tabular}{|c|c|c|c|c|}
			\hline
			\rule{0pt}{15pt}	&$f_{\theta_{1}}(\mathbf{y}_{1})$ &$f_{\theta_1}\left(\mathbf{g}_{1,2}\mathbf{x}_2\right)$ & $f_{\theta_{2}}(\mathbf{y}_{2})$ & $f_{\theta_{2}}\left(\mathbf{g}_{2,1}\mathbf{x}_{1}\right)$ \\ \hline
			\rule{0pt}{15pt}  	PSNR(dB) & 25.72
			&7.34 & 25.68  &7.20 \\ \hline
			\rule{0pt}{15pt}  	MS-SSIM &0.928&0.066& 0.928  &0.057\\ \hline
		\end{tabular}
		\label{rr2}
	}
\end{table}

\begin{figure}[!t]
	\centering
	\begin{subfigure}[The input images of User 1 and User 2]{%
			\includegraphics[width=4cm]{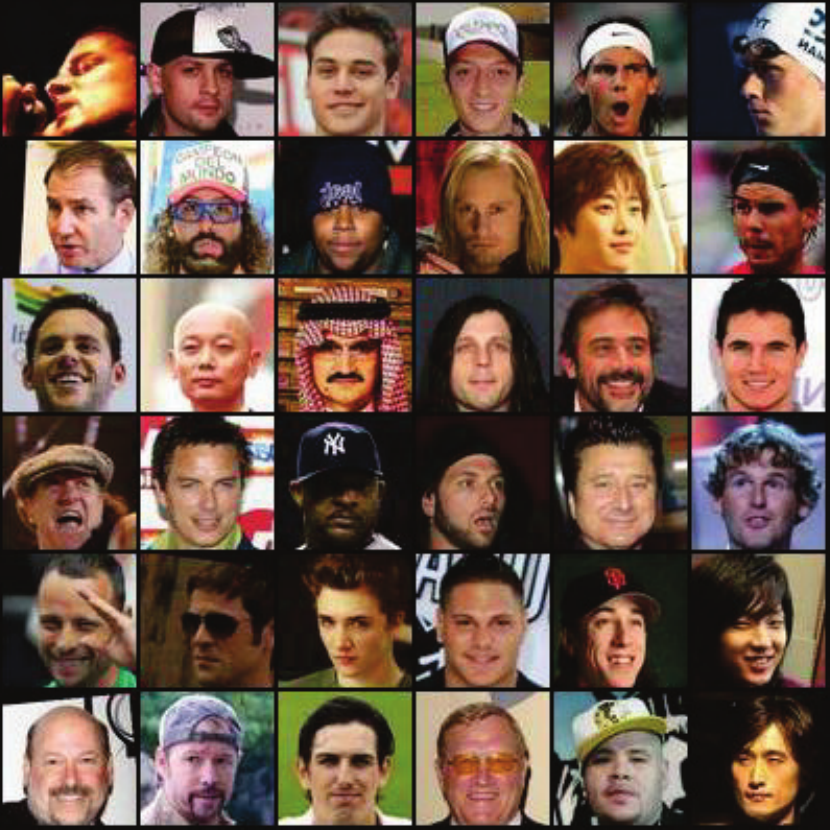}
			\label{input}}
	\end{subfigure}
	\qquad
	\begin{subfigure}[The decoded  images of User 1  ${f_{{\theta _1}}}\left( {{{\bf{y}}_1}} \right)$]{%
			\includegraphics[width=4cm]{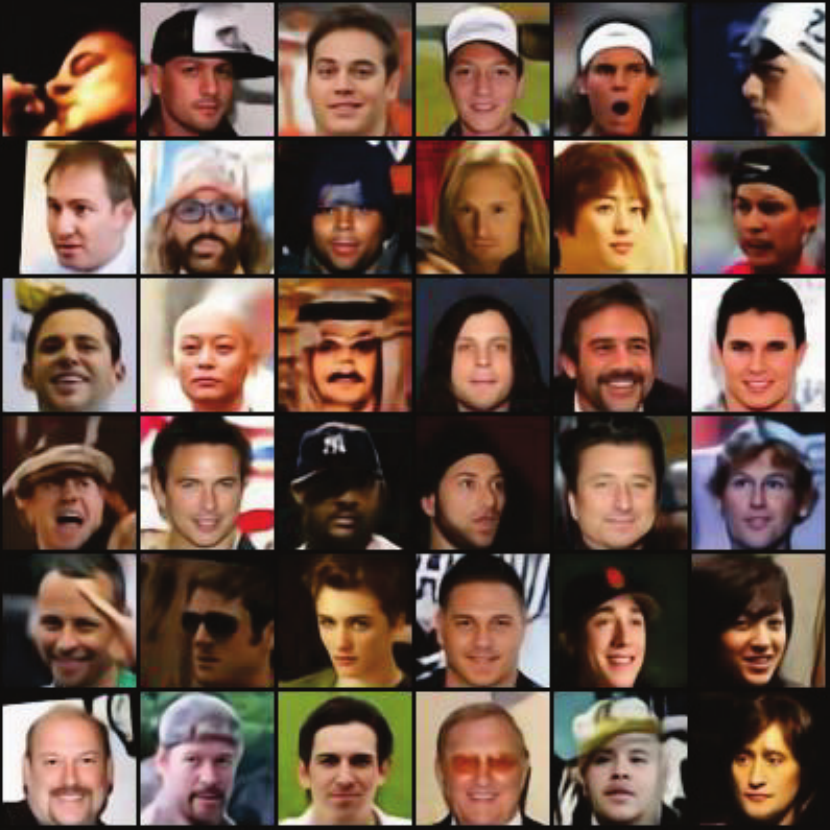}
			\label{1d1}}
	\end{subfigure}
	\begin{subfigure}[The decoded  images of User 2  ${f_{{\theta _2}}}\left( {{{\bf{y}}_2}} \right)$]{%
			\includegraphics[width=4cm]{photo/Ix1_re_5dB.pdf}
			\label{2d2}}
	\end{subfigure}
	\begin{subfigure}[User $1$ decoded  images of User 2 ${f_{{\theta _1}}}\left( {{\bf{g}}_1^H{{\bf{x}}_2}} \right)$]{%
			\includegraphics[width=4cm]{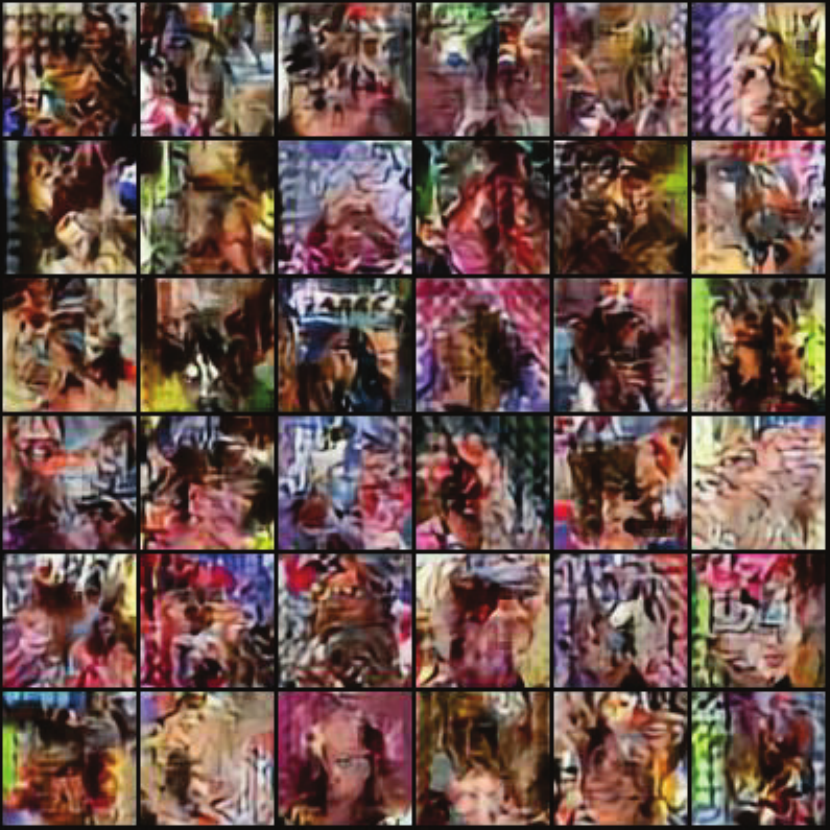}
			\label{1d2}}
	\end{subfigure}
	\begin{subfigure}[User $2$ decoded  images of User 1 ${f_{{\theta _2}}}\left( {{\bf{g}}_2^H{{\bf{x}}_1}} \right)$]{%
			\includegraphics[width=4cm]{photo/samedata_5psnrx2_recon_celebatest.pdf}
			\label{2d1}}
	\end{subfigure}
	\caption{The semantic decoded images of User $1$ and $2$ with the same inputs.}
	\label{same input}
\end{figure}

\begin{figure}[t]
	\centering
	\begin{subfigure}[Training   ${\rm{SNR}=5dB}$]{%
			\includegraphics[width=6cm]{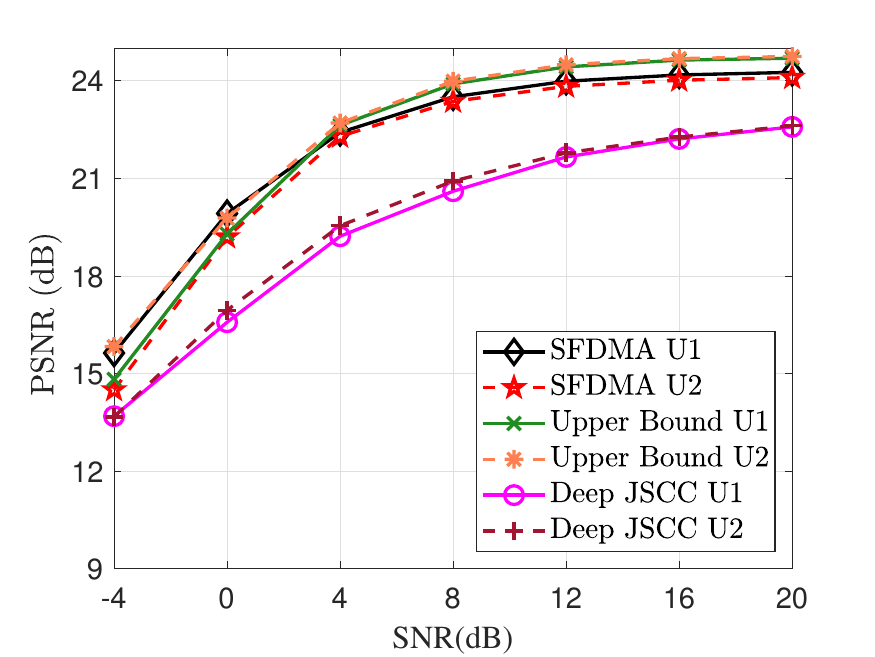}
			\label{last11}}
	\end{subfigure}
	\begin{subfigure}[Training   ${\rm{SNR}=5dB}$]{%
			\includegraphics[width=6cm]{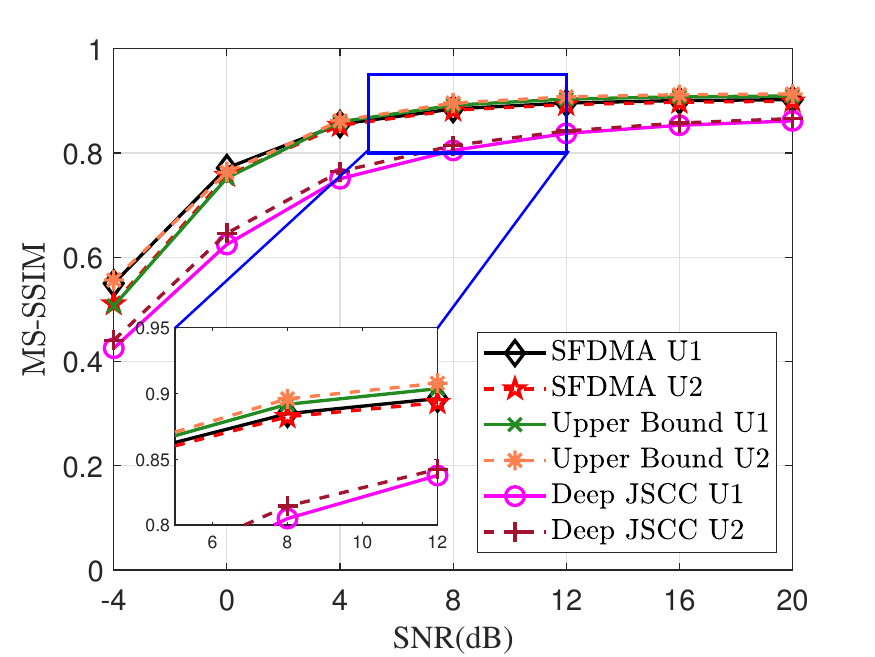}
			\label{last12}}
	\end{subfigure}
	\caption{PSNR and MS-SSIM of image reconstruction of the semantic BC network over Rayleigh channels.}
	\label{PSNR2}
\end{figure}

\subsection{Performance Evaluation for Image Reconstruction Tasks}

To evaluate the experimental results of image reconstruction, PSNR and MSSIM are employed to measure image reconstruction quality.

\setcounter{figure}{11}
\begin{figure}[!htb]
	\centering
	\begin{subfigure}[CDF of classification accuracy $\phi_1$ for User 1]{%
			\includegraphics[width=6cm]{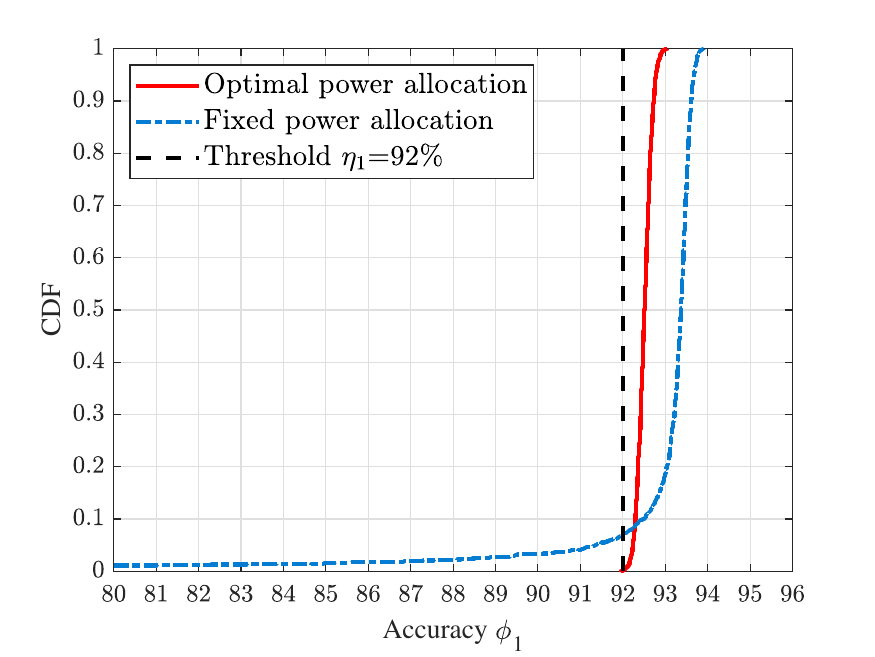}
			\label{cdf_u1}}
	\end{subfigure}
	\begin{subfigure}[CDF of classification accuracy $\phi_2$ for User 2]{%
			\includegraphics[width=6cm]{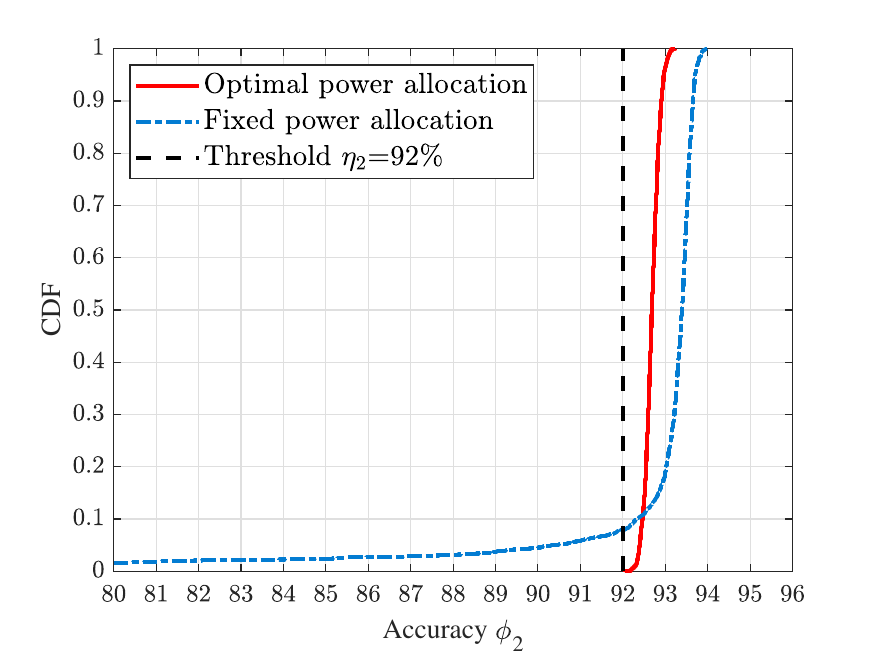}
			\label{cdf_u2}}
	\end{subfigure}
	\caption{CDFs of classification accuracies of  User 1 and User 2 with  different power allocation methods.}
	\label{cdf}
\end{figure}

\begin{figure}[!htb]
	\centering
	\raggedbottom
	\begin{subfigure}[CDF of MS-SSIM $\phi_1$ for User 1]{%
			\includegraphics[width=6cm]{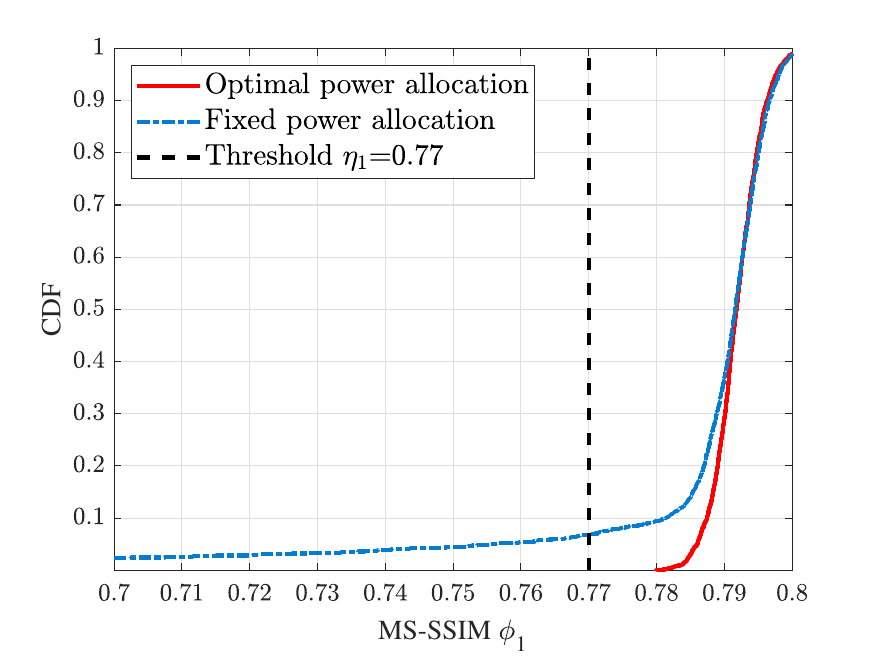}
			\label{cdf_recon_u1}}
	\end{subfigure}
	\begin{subfigure}[CDF of MS-SSIM $\phi_2$ for User 2]{%
			\includegraphics[width=6cm]{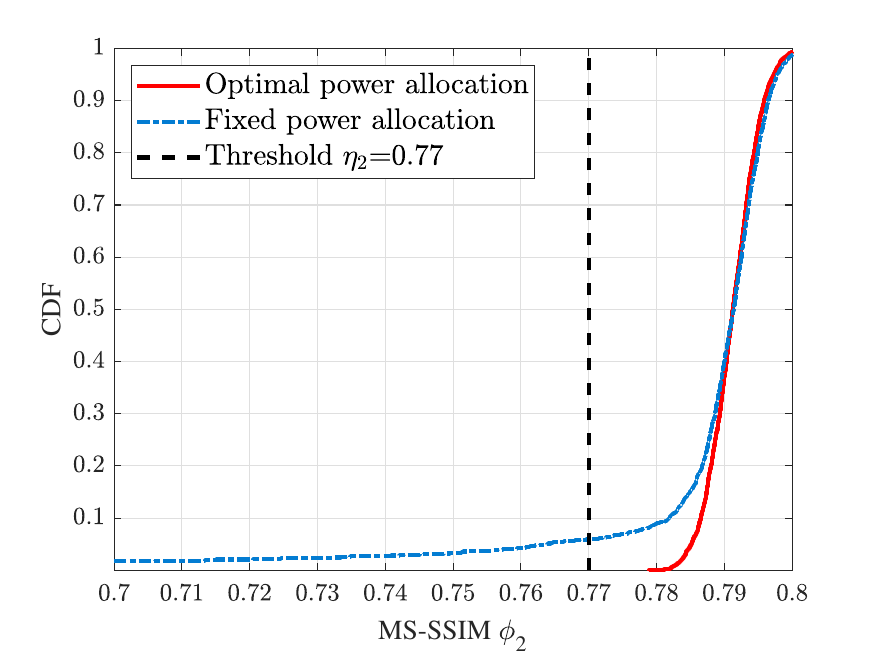}
			\label{cdf_recon_u2}}
	\end{subfigure}
	\caption{CDFs of MS-SSIM of  User 1 and User 2 with  different power allocation methods.}
	\label{cdf_recon}
\end{figure}

 Fig.~\ref{fig:re1} and \ref{fig:re3}   illustrate two-dimensional t-SNE embedding of
  the  semantic features     of the proposed SFMDA BC schemes with two users and three users respectively,   where the  quantization bits ${q_{{\rm{bit}}}} = 4096$ bits and the training SNR = 5dB. Fig.~\ref{fig:re1} and \ref{fig:re3} show  that the   semantic features $\left\{ {{{\bf{x}}_i}} \right\}_{i = 1}^N$  are approximately separated, which verifies the approximate orthogonality property of semantic features of the proposed SFDMA scheme.
For the two users scenario,  the  PSNR of the image reconstruction    of User $1$ and $2$ are
 $21.00$dB and $21.45$dB respectively, and the     MS-SSIM of the image reconstruction    of User $1$ and $2$ are
  $0.802$ and    $0.806$, respectively.
 Moreover, for the three users scenario,  the  PSNR of the image reconstruction    of User $1$,  $2$ and $3$ are
 $24.76$dB, $24.93$dB and $25.00$dB respectively, and the     MS-SSIM of the image reconstruction    of User $1$,  $2$ and $3$ are
  $0.907$, $0.909$ and    $0.910$, respectively.


\setcounter{figure}{10}
\begin{figure*}[!bt]
	\centering
	\begin{subfigure}
		[ The input images for User $1$]{%
			\includegraphics[width=5cm]{photo/Ix1.pdf}
			\label{input_U1}}
	\end{subfigure}
	\qquad
	\begin{subfigure}[The   decoded  image of SFDMA  for User $1$ with training and testing SNR=$-5$dB]{%
			\includegraphics[width=5cm]{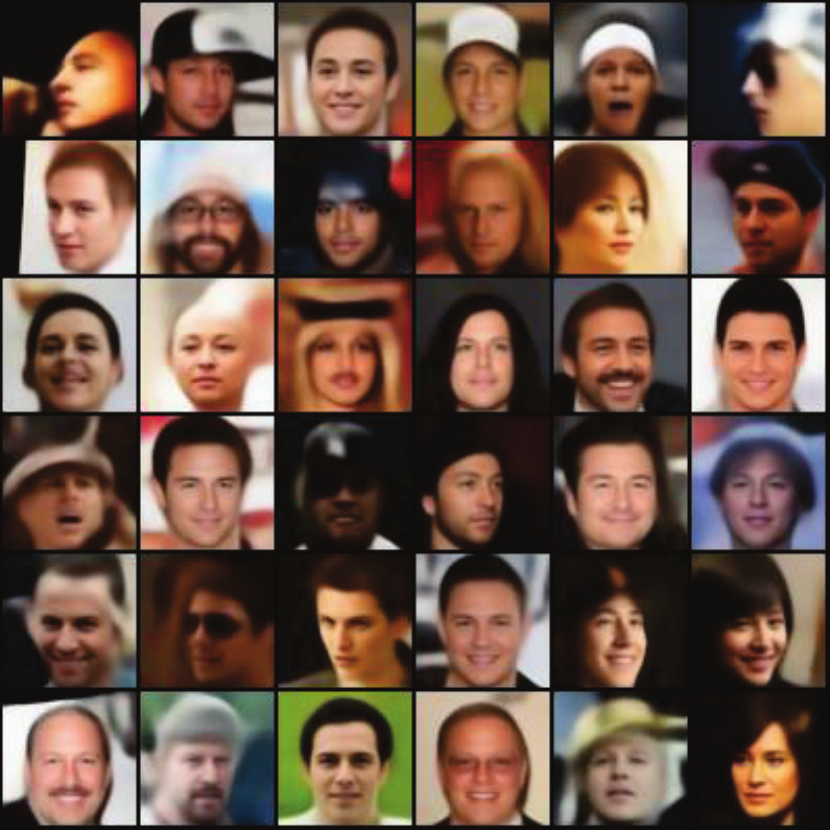}
			\label{U1_05}}
	\end{subfigure}
	\qquad
	\begin{subfigure}[The   decoded  image of SFDMA  for User $1$ with training and testing SNR=$5$dB]{%
			\includegraphics[width=5cm]{photo/Ix1_re_5dB.pdf}
			\label{U1_5}}
	\end{subfigure}
	\begin{subfigure}[The input images for User $2$]{%
			\includegraphics[width=5cm]{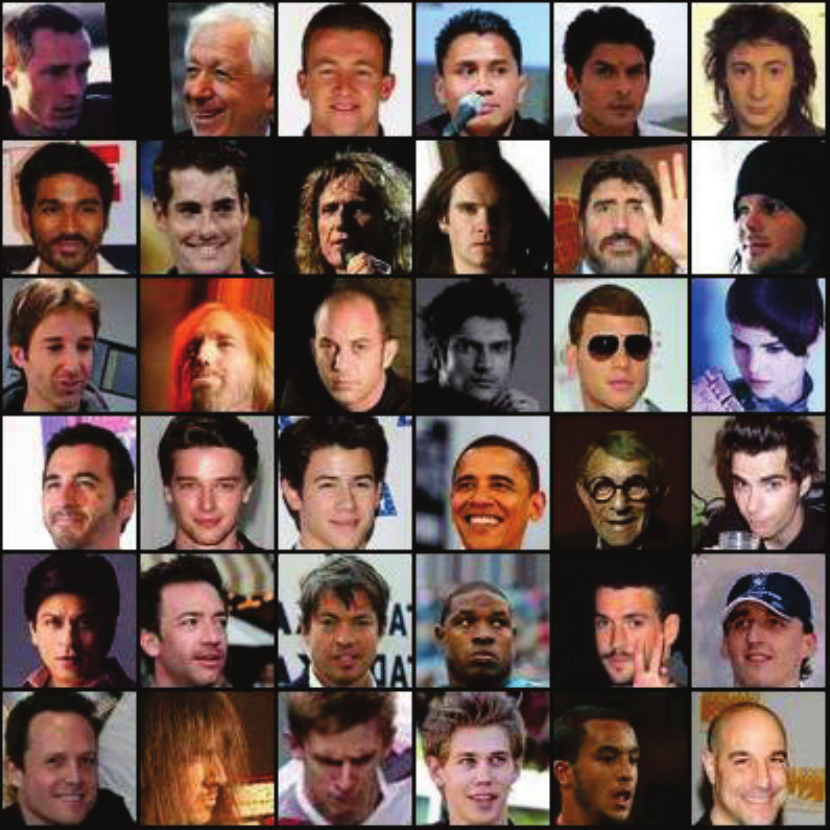}
			\label{input_U2}}
	\end{subfigure}
	\qquad
	\begin{subfigure}[The   decoded  image of SFDMA  for User $2$ with training and testing SNR=$-5$dB]{%
			\includegraphics[width=5cm]{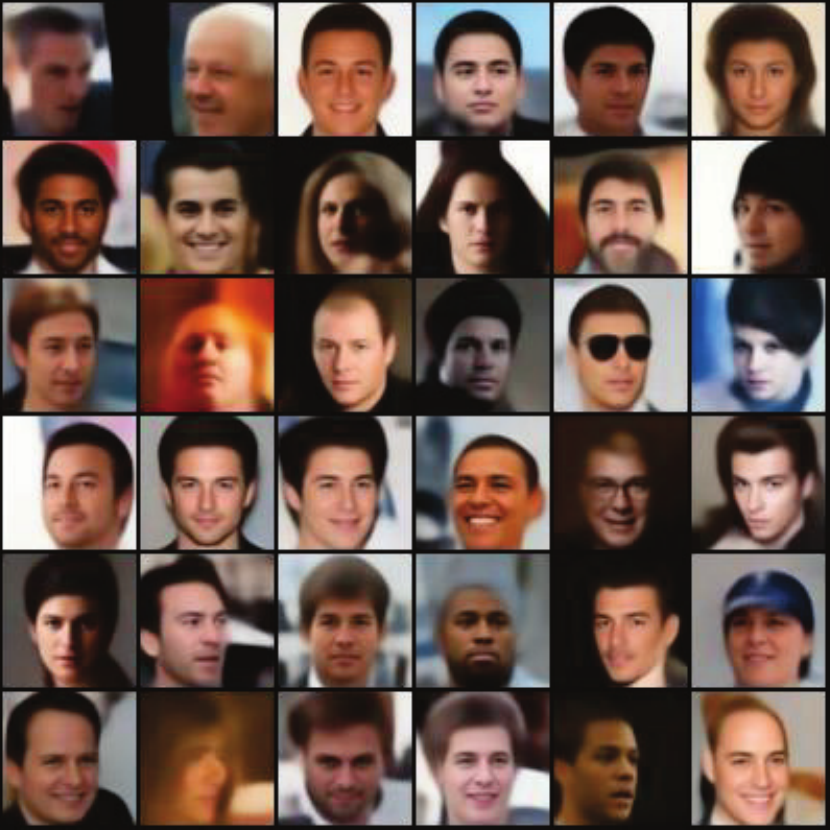}\label{U2_05}}
	\end{subfigure}
	\qquad
	\begin{subfigure}[The   decoded  image of SFDMA  for User $2$ with training and testing SNR=$5$dB]{%
			\includegraphics[width=5cm]{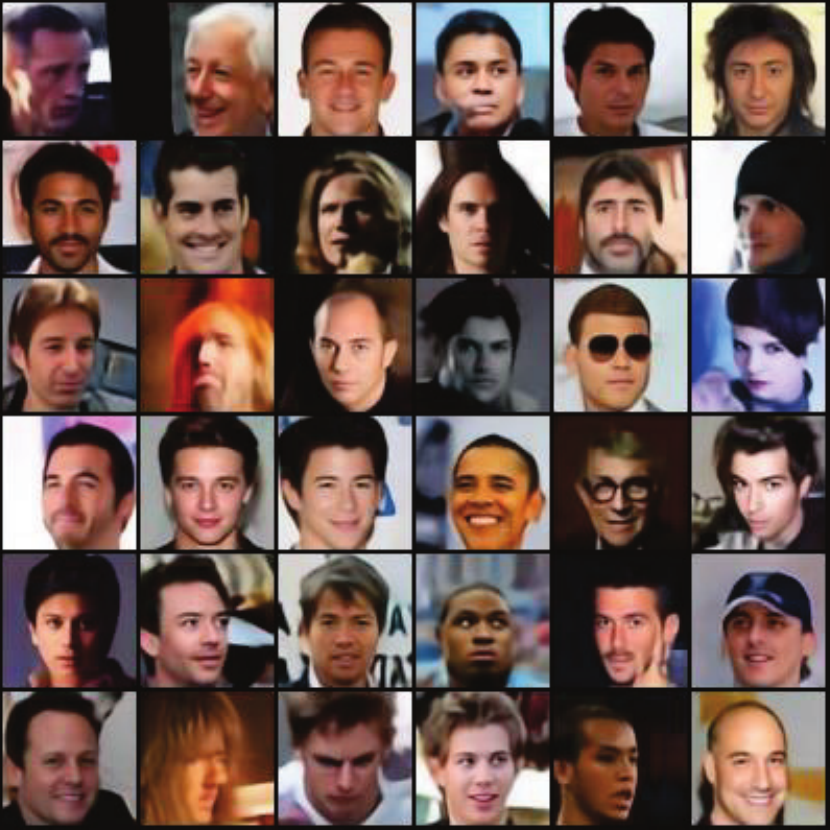}\label{U2_5}}
		\caption{ The decoded  images of User $1$ and $2$ of the proposed SFDMA BC network.}
		\label{fig444}
	\end{subfigure}
\end{figure*}

  Table~\ref{tr} illustrates the inner product and the angle among the semantic features ${\bf{x}}_1$, ${\bf{x}}_2$ and ${\bf{x}}_3$  with the same input images.
Table~\ref{tr} shows that even with the same input images, the inner products among the semantic features of different users are $-1.0  \times {10^{ - 3}}$, $7.0 \times {10^{ - 3}}$ and $ -2.8 \times {10^{ - 4}}$,which approach zero. Moreover, the corresponding angles are $90.06^ \circ$, $89.60^ \circ$ and $90.02^ \circ$, which approach ${\rm{90}}^ \circ $.
  The results in Table~\ref{tr} verify    the semantic features of multi-user
are approximately orthogonal  for the proposed SFDMA scheme.

Table~\ref{rr2} further illustrates orthogonality through the image reconstruction performance using the same input images, which is shown in Fig.~\ref{same input}.
As shown in  Table~\ref{rr2},    with the same inputs,   the PSNRs of image reconstruction of User 1 and 2 decoding their   intended signals  are  $25.72$ dB and $25.68$ dB, respectively, and the corresponding MS-SSIMs are $0.928$ and $0.928$, respectively, the  decoded images are shown in
 Fig.~\ref{1d1} and \ref{2d2}, respectively.

 Moreover,  the PSNR and MS-SSIM of image reconstruction of User 1 decoding  User 2's signal  are $7.34$dB and $0.066$, respectively, and  the  decoded images are shown in Fig.~\ref{1d2}.
  The   PSNR and MS-SSIM of image reconstruction of User 2 decoding  User 1's signal  are $7.20$dB and $0.057$, respectively, and  the  decoded images are shown in Fig.~\ref{2d1}.
    The results in Table~\ref{rr2}  demonstrate that each user can accurately decode their intended signals, and  cannot effectively decode the semantic information of other users, which also
  verify    the semantic features of multi-user
are approximately orthogonal  for the proposed SFDMA scheme.
  Moreover, Table~\ref{table1} also demonstrates
 the proposed SFDMA can   protect the  semantic information from being decoded by other users.

Fig.~\ref{last11}
illustrates   PSNRs   of  Deep JSCC,  Upper Bound and the proposed SFDMA for the semantic BC network versus  SNR over  Rayleigh channel with the training   ${\rm{SNR}=0dB}$.
Fig.~\ref{last11}   show  that, as SNR
increases,    PSNRs of three schemes   increase  rapidly at first, and then slowly
increases until it reaches the maximum value.
Moreover, PSNRs of the proposed SFDMA scheme are higher than that of Deep JSCC, and approach to  that of the Upper bound.
Fig.~\ref{last12}
illustrates   MS-SSIM   of  Deep JSCC,  Upper Bound and the proposed SFDMA for the semantic BC network versus  SNR over  Rayleigh channel with the training   SNR=$5$dB.
Similar  to Fig.~\ref{last12}, as SNR
increases,    MS-SSIMs of three schemes   increase  rapidly at first, and then slowly
increases until it reaches the maximum value, and the PSNRs of the proposed SFDMA scheme are higher than that of Deep JSCC, and approach to  that of the Upper bound.
Fig.~\ref{last11} and Fig.~\ref{last12}   verify  the proposed SFDMA scheme can effectively eliminate multi-user interference  and improve image  reconstruction tasks performance.

%
Furthermore, Fig.~\ref{fig444} illustrates the decoded  images of User $1$ and $2$ of the proposed SFDMA BC network. Fig.~\ref{input_U1} and \ref{input_U2} show the input images for User $1$ and $2$, respectively. Fig.~\ref{U1_05} and \ref{U1_5} show the decoded  image of SFDMA  for User $1$ with training and testing SNR=$-5$dB and SNR=$5$dB, respectively.  Fig.~\ref{U2_05} and \ref{U2_5} show the decoded  image of SFDMA  for User $2$ with training and testing SNR=$-5$dB and SNR=$5$dB, respectively. Fig.~\ref{fig444} demonstrates that   the proposed SFDMA can effectively support images transmission for  multi-user BC networks.

\subsection{ Performance of power allocation schemes}

Fig.~\ref{cdf} illustrates the cumulative distribution functions
(CDFs) of classification accuracy of the fixed
power allocation method and  the proposed  power allocation
method, where  the classification accuracy thresholds $\eta_1$ and $\eta_2$ are both $90\%$.
Fig.~\ref{cdf} shows the classification accuracy of both two users under the optimal power allocation method is higher than the $90\%$ threshold, which satisfies the inference tasks requirement, whereas the fixed power allocation method exhibits a probability of the classification accuracy falling below the threshold.

Fig.~\ref{cdf_recon} illustrates the CDFs of MS-SSIM for the image reconstruction task. With MS-SSIM thresholds set at 0.77, the proposed method consistently maintains MS-SSIM consistently above the threshold. Additionally, its peak performance is comparable to that of the fixed power method, which often falls below the threshold. This demonstrates the superior performance of the proposed approach.
Fig.~\ref{cdf} and Fig.~\ref{cdf_recon} verifies  that the
propose power allocation  method can effectively guarantee the
QoS of semantic BC network.
\section{Conclusions}

In this paper, we proposed a SFDMA scheme for the multi-user BC network with inference and image reconstruction tasks, in which the information of multiple users are encoded  in distinguishable feature subspaces. The semantic encoded features of multiple users are simultaneously transmitted over the shared communication channel.
Specifically, for our proposed SFDMA scheme, the encoded semantic features of  multiple users are approximately   orthogonal to each other, hence can be simultaneously transmitted in
the same time-frequency resource.   For   inference tasks,  we developed  a  RIB based SFDMA   BC network   to 	  achieve a tradeoff between   inference performance,
data compression and multi-user interference.
	   For   image  reconstruction tasks,  we designed   a Swin Transformer based   SFDMA   BC network, in which  multi-user interference is  significantly reduced, and each user can efficiently decode the intended images.
Moreover, our proposed SFDMA scheme can protect the privacy of users' semantic information  from  being decoded  by other users.
   Furthermore, we  proposed the   ABG formula  to fit   the relationship between task performance and SINR.
    Based on the ABG function,  we proposed  an optimal power allocation   method  for semantic BC networks with  inference and reconstruction tasks. Our proposed  power allocation method   can   effectively guarantee the  performance requirements of semantic users in random fading channels.

\bibliographystyle{IEEEtran}

\bibliography{reference}

\end{document}